\documentclass[aps,prx,twocolumn,nofootinbib]{revtex4}
\usepackage{bm, graphicx, amsmath, amssymb}
\usepackage{bbm}
\usepackage[mathscr]{eucal}
\usepackage{epsfig,graphics,graphicx}
\usepackage{xcolor}
\usepackage{lipsum}
\usepackage{enumitem}
\usepackage{ytableau}

\usepackage[T1]{fontenc}

\usepackage{hyperref}
\hypersetup{citecolor=blue}
\hypersetup{colorlinks=true}

\newcommand{\R}{\mathbb{R}}

\newcommand{\grp}[1]{\mathsf{#1}}
\newcommand{\spc}[1]{\mathcal{#1}}

\def\d{{\rm d}}

\def\>{\rangle}
\def\<{\langle}
\def\kk{\>\!\>}
\def\bb{\<\!\<}
\newcommand{\st}[1]{\mathbf{#1}}

\newcommand{\Vac}[1]{\mathrm{Vac}}

\newcommand{\map}[1]{\mathcal{#1}}
\newcommand{\Tr}{\operatorname{\sf Tr}}

\newcommand{\id}{I}
\newcommand{\tr}{{\rm tr}\,}
\newcommand{\ket}[1]{|{#1}\rangle}
\newcommand{\bra}[1]{\langle{#1}|}

\newcommand{\ketbrad}[1]{|{#1}\rangle\!\langle{#1}|}

\newcommand{\sfr}[2]{\mbox{\small$#1\over#2$}}

\usepackage{mathtools}

\begin{document}

\title{Quantum learning  with a single-atom  sensor}

\author{Mo Yin$^{1}$, Emilio Bagan$^{2,3}$ and Giulio Chiribella$^{3,4,5}$}

\affiliation{
$^1$  Thrust of Artificial Intelligence, Information Hub, The Hong Kong University of Science and Technology (Guangzhou).\\
$^{2}$F\'{i}sica Te\`{o}rica: Informaci\'{o} i Fen\`{o}mens Qu\`antics, Departament de F\'{\i}sica, Universitat Aut\`{o}noma de Barcelona, 08193 Bellaterra (Barcelona), Spain\\
$^{3}$ QICI Quantum Information and Computation Initiative, School of Computing and Data Science,  The University of Hong Kong, Pokfulam Road, Hong Kong\\
$^4$ Quantum Group, Department of Computer Science, University of Oxford, Wolfson Building, Parks Road, Oxford, OX1 3QD, United Kingdom\\
$^5$ Perimeter Institute for Theoretical Physics, 31 Caroline Street North, Waterloo, Ontario, Canada}

\begin{abstract}
The  ability to  gather information and to act  upon it is at the core  of every learning agent.       
 But what is the impact of quantum mechanics on an agent's ability to sense external inputs and to  translate them   into actions? 
   Here we  address the question for a prototype task of learning agency at the quantum scale: rotating a single  spin based on information gathered by a single atom.
   We determine the ultimate performance limit for  this task, revealing a fundamental tradeoff between entanglement at the sensing stage and coherence at the action stage: if the single-atom sensor is not entangled with the quantum system serving as the agent's internal memory,  then  the best learning strategy requires a coherent transfer of quantum information from the sensor to the system  that controls the agent's actions. 
    In contrast, if the sensor is initially entangled with the agent's memory,  then the transfer of quantum information is no longer necessary.   Our results indicate that the quantum properties of the sensor radically affect the optimal way to convert external stimuli into actions,  revealing a link   between quantum sensing and the behavior of quantum agents.   
    \end{abstract}

\maketitle

A distinctive trait of learning agents  is their ability to gather information and to use it to perform actions.     Abstractly, a learning agent can be thought as a machine consisting  of three main components: sensors for data collection,  an internal memory for processing and storage, and  actuators that  enable the machine  to act  upon its environment \cite{russel2013artificial}. 
While current applications of machine learning mostly involve macroscopic  machines governed by classical physics, the growing development of quantum technologies suggests the potential for engineering learning machines that operate at the quantum scale \cite{schuld2015introduction,biamonte2017quantum, dunjko2018machine},   
 leading to  faster solutions to classical learning problems \cite{aimeur2006machine, harrow2009quantum, rebentrost2014quantum, ronnow2014defining, wiebe2014quantum, dunjko2016quantum, amin2018quantum}, to a reduction of memory and energy resources  \cite{elliott2022quantum,thompson2025energetic}, and to the exploration of  new, genuinely quantum learning tasks,  such as
 quantum state classification \cite{sasaki2001quantum, sasaki2002quantum, guctua2010quantum, sentis2012quantum}, unitary gate learning \cite{bisio2010optimal,sedlak2018optimal} and emulation \cite{marvian2025efficient},  measurement learning \cite{bisio2011quantum, sentis2012quantum}, and learning of isometric channels~\cite{yoshida2026quantum}. 

Most works in quantum machine learning assume that the information gathered by the machine is {\em fungible},~{\em i.e.}  independent of the physical means of encoding. However, learning machines embedded  in specific physical environments often have to deal with non-fungible information \cite{bartlett2007reference}.  For instance, a machine operating in an unknown magnetic field may need to gather information about the field's orientation relative to its proper frame, necessitating a sensor specifically coupled to the magnetic field. The physics of the sensor can significantly impact the quality of the gathered information, especially in the case of quantum sensors, which can have radically different  responses to external parameters, depending on quantum resources such as coherence, entanglement, and asymmetry \cite{degen2017quantum,pirandola2018advances,marvian2013theory,ahmadi2013wigner,marvian2014extending}. Nevertheless, the interplay between  quantum sensors and other components of a learning machine, such as internal memory and actuators, remains largely unexplored.  
Little is known about the best way to convert sensor data into actions:  Is it necessary to transfer quantum information from the sensors to the actuators, or is it sufficient to measure the sensors and store  the outcomes in a classical digital memory? Does the performance improve if the sensors are entangled with other components of the learning machine? 


In this paper we address the extreme scenario where 
   the sensor  consists of  a single atom and the actions are performed on a single quantum spin. The task is to rotate the direction of the spin, by an initially unknown amount and about an initially unknown direction, which are learnt by probing  the action of the rotation  on the atom. Classical algorithms for learning rotations are well-studied  \cite{arora2009learning} due to their applications to  computer vision \cite{eggert1997estimating}, robotics \cite{makadia2006rotation}, and navigation \cite{farrell1966least}. In the quantum domain, protocols for estimating rotational parameters  have been developed in a series of works \cite{gisin1999spin,peres2001entangled,peres2001transmission,bagan2001aligning,acin2001optimal,lindner2003elliptic,bagan2004entanglement,chiribella2004efficient,bagan2004quantum,chiribella2005optimal,hayashi2006parallel,kolenderski2008optimal}. Initially motivated by  the  problem of aligning quantum reference frames \cite{bartlett2007reference,skotiniotis2012alignment}, the optimal estimation of rotations  has  since found applications in quantum communication \cite{bartlett2009quantum}, quantum error correction  \cite{faist2020continuous,woods2020continuous,hayden2021error,yang2022optimal} and to the design of  optimal programmable  gates \cite{yang2020optimal}.   


Estimation is just one possible strategy for learning how to perform an unknown rotation.   
This strategy turns out to be optimal  when  the sensor is a pair of quantum systems  prepared in the optimal quantum state  \cite{bisio2010optimal}, but the situation is generally different  when the sensor's  state  is of a restricted form  \cite{marvian2025efficient}  or when some prior information about the rotations is available \cite{mo2019quantum}. Here, we consider the fundamental problem of learning completely unknown rotations using  general states  of a  single hydrogen atom,   a setting   originally  envisaged in the pioneering work by Asher Peres and coworkers \cite{peres2001transmission,lindner2003elliptic}. In these settings, finding the best learning strategy is a hard  problem, involving two coupled optimizations:  optimizing the atom's state used to capture information about the unknown rotation, and optimizing the process that converts the atom's state   into an action performed on the target spin. 
 The combination of these two optimization problems is not a semidefinite program, and its numerical solution  is not viable when the dimension of the quantum state space  becomes large.

\begin{figure}[h]
    \includegraphics[width=0.9\linewidth]{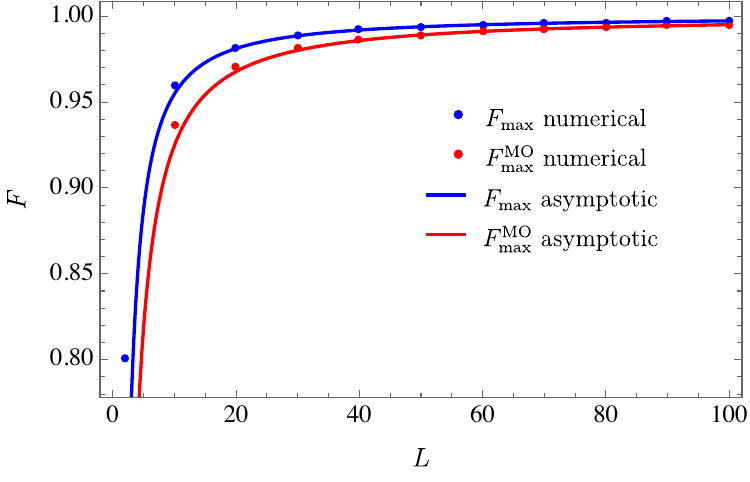}
    \caption{\emph{Ultimate quantum limit for learning rotations from a single atom.} The figure shows the fidelity between the output produced by the learning machine and the output of the given rotation, for the optimal quantum learning machine (blue) and the optimal machine based on estimation (red). The  solid curves  represent the asymptotic expressions  (\ref{Fopt})  (blue) and (\ref{FMO}) (red), respectively, while   dots are the results of numerical optimization.    }
    \label{fig:Fopt}
\end{figure}

In spite of the technical challenges, here we provide a complete solution, identifying the in-principle ultimate  quantum strategy for learning an arbitrary, initially unknown rotation from a single atom's state.   For large values of the atom's energy, we show that the optimal quantum fidelity  between the machine's operation and the desired rotation can be evaluated analytically up to  fourth order asymptotics, yielding  the expression 
\begin{equation}
F_{\rm max}   = 1-{1\over {6L}}+{\gamma_1\over {6L^{4/3}}}-{4  \gamma_1^2\over {45 L^{5/3}}}+O(L^{-2}),
\label{Fopt}
\end{equation}
where  $L$ is the maximum value of the atom's orbital angular momentum compatible with a given value of the energy,   and $\gamma_1\approx -2.3381$ is the first zero of the Airy function of the first kind.  Remarkably, Eq. (\ref{Fopt}) is an accurate approximation also for small values of $L$, as shown in Fig. \ref{fig:Fopt},    where we compare Eq. (\ref{Fopt}) with numerical evaluation of the exact expression of the fidelity. 

We then show that  the optimal learning machine in the single-atom setting surpasses all  learning machines based on estimation.  Even more generally, we show that  the optimal learning machine  achieves higher fidelity than  any machine that measures  the atom  and operates on the spin based on the measurement result.   To this purpose, we identify the optimal measure-and-operate strategy, show that it coincides with estimation, and derive  the asymptotic expression   
\begin{equation}
F^{\rm MO}_{\rm max}   = 1-{1\over {3L}}+ {2^{2/3}\gamma_1\over {6L^{4/3}}}-{2^{7/3}  \gamma_1^2\over {45 L^{5/3}}} +O(L^{-2}) .
\label{FMO}
\end{equation}
Eq. (\ref{FMO})  generalizes and improves upon previous results on the optimal estimation of rotations \cite{peres2001transmission,bagan2001aligning}, which  only provided the first two orders in the asymptotics. Thanks to the higher order corrections,  Eq. (\ref{FMO}) is a good approximation also for small values of $L$ (cf. Fig.~\ref{fig:Fopt}).   

Comparing Eqs. (\ref{Fopt}) and (\ref{FMO}),  we conclude  that the optimal strategy requires a transfer of quantum information from the sensor to the machine's internal memory, and eventually, to the actuators.  Moreover, Eq. (\ref{FMO}) provides a rigorous benchmark that can be used to certify  genuine quantum transfer of information from memory to actions even in the presence of noise and imperfections. 

 \begin{figure}[h]
    \includegraphics[width=1\linewidth]{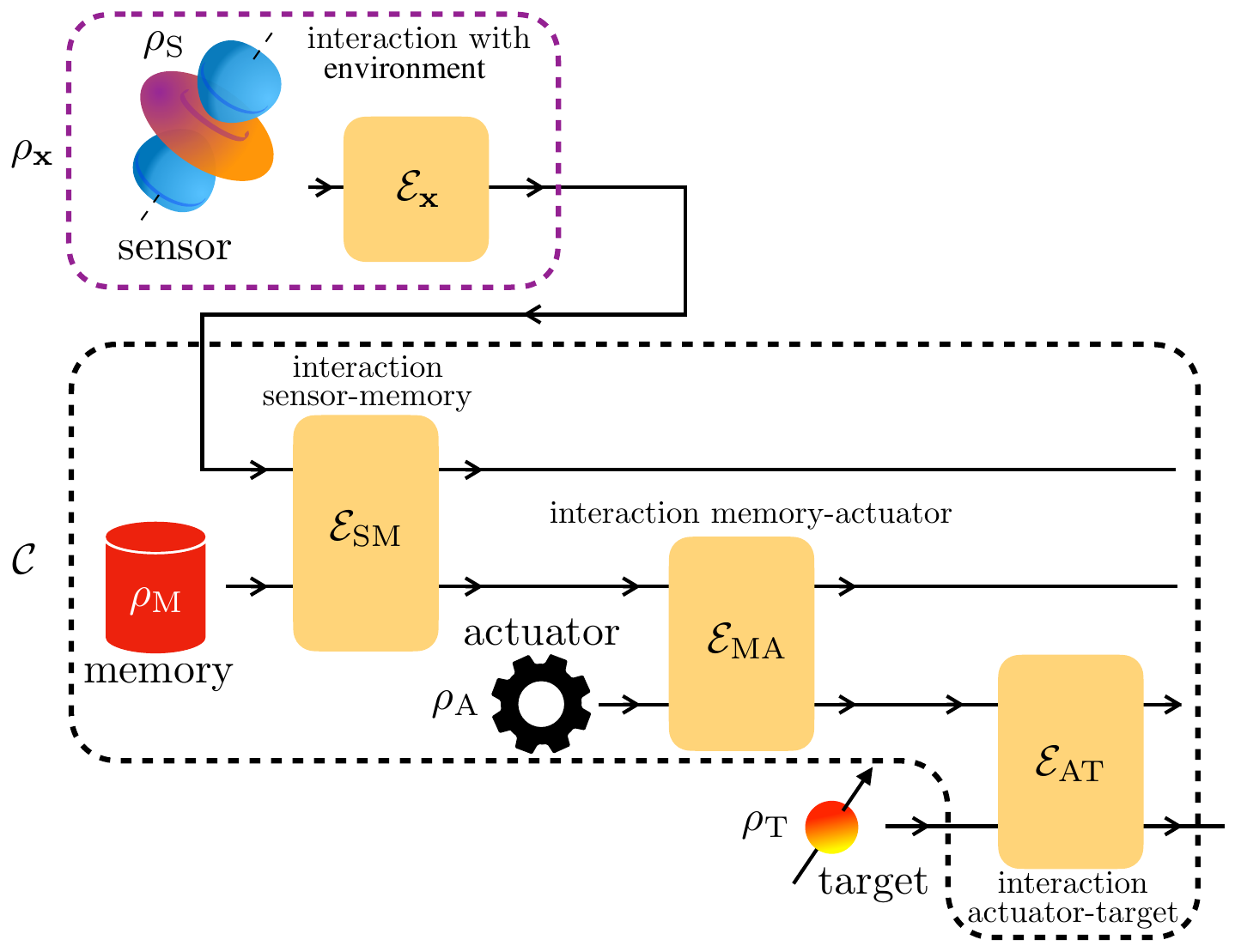}
    \caption{\emph{Learning from a quantum sensor.}    A sensor $\rm S$, initially in the state $\rho_{\rm S}$,  interacts with the environment, gathering information about  a set of parameters $\bf x$.  After the interaction, the sensor ends up in the state  $\rho_{\bf x}$.  It then transfers information to  the internal memory $\rm M$ of the learning machine, initially in the state $\rho_{\rm M}$.    When the machine is required to operate, it accesses its memory by letting it  interact with a quantum system $\rm A$, called the actuator, which controls the machine's operations on a target system $\rm T$.   The overall process  converting sensorial data  into actions on the target system is described by an effective channel $\map C$, with input $\rm S  T$ and output $\rm T$.    }
    \label{fig:machine}
\end{figure}

Finally, we show  that the situation completely changes if the  quantum sensor and the machine's memory are initially entangled. In this case, the  learning fidelity optimized over all possible entangled states has  the Heisenberg scaling  
\begin{align}\label{Fheisenberg}
F_{\max}^*  =  1- \frac{ \pi^2}  {6  L^2}   +  O(  L^{-3})   \, .
\end{align}  
  This scaling is achieved by estimation, as in the optimal unconstrained strategy of Ref. \cite{bisio2010optimal} (see the Supplemental Material \cite{supplemental} for details). Hence,  no quantum information needs to be transferred from the sensor to the machine's actuators. Our findings indicate that the quantum properties of the sensors  affect the internal structure of the optimal learning agent.

{\em Framework.}   Our framework,  illustrated in Fig. \ref{fig:machine}, is inspired by the fully quantum model of learning machines by Dunjko,  Taylor, and  Briegel \cite{dunjko2016quantum}.  A quantum sensor $\rm S$, initially in the state $\rho_{\rm S}$,  interacts with an environment  characterized by some  parameters  $\st x$. As a result of the interaction, $\rm S$   ends  up in the state $\rho_{\st x}  =  \map E_{\st x}  (\rho)$, where  $\map E_{\st x}$ is a suitable quantum channel (trace-preserving completely positive map).  
Then,  $\rm S$  transfers information to  a part of the machine's internal memory,  denoted by $\rm M$.    The transfer of information is  described by another quantum channel $\map E_{\rm SM}$  acting on the composite system $\rm SM$. Finally, the state of the memory is converted into an action performed by the machine on a target system $\rm T$.    Mathematically, the process of converting states of the memory into actions is  described by a quantum supermap \cite{chiribella2008transforming,chiribella2009theoretical}.  Physically,   the conversion from memories to actions is mediated by a  quantum system $\rm A$, called the actuator \cite{dunjko2016quantum}, initially in some state $\rho_{\rm A}$.    The memory and the actuator first interact together through  a quantum channel $\map E_{\rm MA}$, and then the actuator interacts with the target system, implementing a transformation on it. The execution of a transformation on the target system depending on the state of the actuator is described by  a  quantum channel   $\map E_{\rm AT}$.


The performance of the machine is evaluated operationally, by performing a test. Based on the outcome of the test, the machine receives a reward. 
 An example of test consists in preparing  the target system in a random state $|\psi\>$,  unbeknownst to the machine, and measuring the target system with  an observable  $O_{\psi, \bf x}$  after the machine has acted.  The outcome of the measurement is then used to quantify the reward.    On average over all possible  outcomes and over all possible  states, the reward is
 \begin{align}\label{generalF}
F (\st x)    =  \int  \d\psi  \,  \Tr \, [  O_{\psi,  \st x}   \, \map C  (  \rho_{\st x} \otimes |\psi\>\<\psi|)  ]  \, ,
\end{align}  
where $\d \psi$ is the normalized Haar measure,  and $\map C$ is the overall quantum channel describing the transformation from the sensor's state to an action on the target (see the box at the bottom of Fig. \ref{fig:machine}).

The  sensing-action-reward sequence  can be the basic building block of  more complex learning processes where a machine adapts its response  to the environment, as in reinforcement learning   \cite{sutton2018reinforcement}.  Here, however, we  focus on  a single instance of the learning process,  optimizing the reward over all possible sensor states $\rho_{\rm S}$ and over all possible quantum channels $\map C$, in the worst case over~all~possible values of $\st x$.    
 The optimal reward achieved in this setting provides an upper bound to the performance achievable  when the task is not known {\em a priori} and the learning process involves multiple steps.

{\em Learning rotations from a single atom.}   
Hydrogen atoms 
are among the first systems to be studied as quantum reference frames for spatial directions  \cite{peres2001transmission,lindner2003elliptic}. In the basic scenario, directional information is encoded into states of the orbital angular momentum within a given energy shell, so that  the quantum number  of the orbital angular momentum takes a finite set of values $l  \in  \{0, 1,\dots,  L\}$ for some integer $L$. An orbital angular momentum   state in the given energy shell can then be written as
\begin{equation}
\ket\phi_{\rm S}=\sum_{l=0}^L c_l \ket{\phi_l}_{\rm S},
\label{c's}
\end{equation}
where $\ket{\phi_l}_{\rm S}$ are unit  vectors belonging to the spin-$l$ representation of the  group   $\grp{SU} (2)$
and $\sum_{l=0}^L|c_l|^2=1$.  When the atom is used to gather information about an unknown spatial rotation $g$,  it undergoes the unitary gate   
$V_g=\bigoplus_{l=0}^L U^{(l)}_g$,
where $U^{(l)}_g$  is the spin-$l$  irrep of  $\grp{SU}  (2)$.     Hence, the state after the encoding is $\ket{\phi_g}:=V_g\ket{\phi}_{\rm S}$. 

For the  target system, we take a qubit embodied by the spin of a spin-$1/2$ particle. The goal of the learning machine is to perform the rotation  $U_g :=  U_g^{({\frac 12})}$ on the target qubit.     To test the performance, we initialize  the target  in a random  pure state $|\psi\>$ and estimate the resemblance of its final state with the state $U_g |\psi\>$.    
This test amounts to choosing the observable $O_{\psi,g}  =  U_g  |\psi\>\<\psi|  U_g^\dag$,  which gives rise to the gate  fidelity  \cite{horodecki1999general,gilchrist2005distance}
\begin{align}\label{Flearn}
F(g)&=\int \d \psi ~  \Tr \left[O_{\psi,g}   ~  {\mathcal C}(|\phi_g\>\<\phi_g|\!\otimes\!|\psi\>\<\psi|) \right]\,, 
\end{align}
corresponding to Eq. (\ref{generalF}) in the specific scenario considered here. 


{\em Ultimate quantum limit.} We now optimize the  fidelity (\ref{Flearn})  in the worst case over all possible rotations $g$. For this purpose, it is convenient to introduce the notations   $|\tilde \phi\>  :  = {\rm e}^{-i\pi L_y}  |\overline\phi\> $, where the bar stands for complex conjugation, and  $\widetilde C:=\left({\rm e}^{-i\pi L_y}\!\otimes\! \id_{\rm T_{out}}\!\otimes\!  \id_{\rm T_{in}}\right) C \left({\rm e}^{i\pi L_y}\!\otimes\! \id_{\rm T_{out}}\!\otimes\! \id_{\rm T_{in}}\right)$, where $L_y$ is the $y$ component of the atom's angular momentum operator, and  $C$ is the Choi operator of the channel $\map C$ \cite{choi1975completely},  acting on the tensor product Hilbert space $\spc H_{\rm S}  \otimes \spc H_{\rm T_{\rm out}}  \otimes \spc H_{\rm T_{\rm in}}$, where ${\rm T_{\rm in}}$ and ${\rm T_{\rm out}}$ denote  the target's system before and after the action of  channel $\map C$, respectively.     With this notation, the  fidelity   can be rewritten as   
\begin{equation}
F(g)   =\frac 13  +  \frac 16    \bra{\tilde\phi_g} \, \!\bb U_g|  \widetilde C \ket{\tilde\phi_g} \, |U_g\kk   \, ,
\label{Fe tilde C}
\end{equation}
with $   |U_g\kk  :  =  (  U_g \otimes \id) |\id\,\kk_{\rm T_{out}  T_{in}}$ and $  |\id\,\kk_{\rm T_{out}  T_{in}}  =  |\sfr12,\sfr12\>_{\rm T_{out}}\otimes |\sfr12,\sfr12\>_{\rm T_{in}}  +  |\sfr12,-\sfr12\>_{\rm T_{out}}\otimes |\sfr12,-\sfr12\>_{\rm T_{in}}$.
 Now, the symmetry of the problem guarantees that the optimal operator $ \widetilde C$ maximizing the worst-case fidelity can be chosen without loss of generality to commute with the representation $ \{V_g \otimes U_g \otimes \id\}_{g\in  \grp{SU}  (2)}$, and therefore to be of the block diagonal form
\begin{align} 
\widetilde C=\bigoplus_{j=\frac12}^{L+  \frac 12} \id^{(j)}\!\otimes\! B^{(j)} \,,
\label{tilde C block}
\end{align} 
where $\id^{(j)}$ is the identity operator on a $(2j+1)$-di\-men\-sion\-al space supporting the spin-$j$ irrep of $\grp{SU}(2)$ and  $B^{(j)}$ is an operator on the space ${\mathbb C}^{\nu_j}\!\otimes {\mathscr H}_{\rm T_{\rm in}}$, where~$\nu_j$ is the multiplicity of the spin-$j$ representation in the tensor product $V_g\otimes U_g$.  Since we have  $\nu_{L+1/2}=1$, and $\nu_j=2$ for \mbox{$j <  L+  1/2$}, it follows that $B^{(L+1/2)}$ is a  $2\times2$ matrix, while  $B^{(j)}$ is a~$4\times4$ matrix for all $j <   L+1/2$.    For the latter we use the notation
\begin{equation}
B^{(j)}=\kern-1.5em\sum_{\sigma,\sigma'   \in  \left\{  -\frac 12, \frac 12 \right\}}\kern-1em\ket{j,\sigma}\bra{j,\sigma'}\!\otimes\!B^{(j)}_{\sigma\sigma'},
\label{B sigma}
\end{equation}
where the states $\{\ket{j,\sigma}\}_{\sigma=\pm1/2}$  are a basis for the multiplicity space for the spin-$j$ irrep, and each  $B^{(j)}_{\sigma\sigma'}$ is a 2-by-2 matrix.

%

In the Supplemental Material \cite{supplemental}, we show that the joint   optimization of the state $|\phi\>$  and the channel $\map C$  yields the achievable bound
\begin{multline}
F \le  \frac 13 +  \frac 1{6}\max_{\{c_l\}}  \left(\sum_{l=2}^L c_l^2{4l+3\over2l+2}\right.\\
\left.+2\sum_{l=2}^L\!c_lc_{l-1}\sqrt{{2l\!-\!1\over2l}}
\!+\! {7\over4} c_1^2
\!+\!\sqrt2 c_1c_0\!+\!c_0^2\!
\right).  \label{optimalF}
\end{multline}
where the maximum is over all unit vectors $\{c_l\}_{l=0}^L$.   The bound is achieved by setting  $\ket{\phi_j}=\ket{j,j}$ for every $j$ in Eq. (\ref{c's}),  and by choosing the matrices
{\em
\begin{align}
B^{(j)}_{{1\over2}{1\over2}}&=\begin{cases}\displaystyle
{2j\over2j+1}\ketbrad{\sfr12 ,\sfr12},\ & j>{1\over2},\\[1em]
\displaystyle{1\over2} \id_{j}, &j={1\over2},
\end{cases}
\\[.5em]
B^{(j)}_{-{1\over2}-{1\over2}}&={2j+2\over2j+1}\ketbrad{\sfr12 ,-\sfr12},\\[.5em]
B^{(j)}_{{1\over2}-{1\over2}}& = B^{(j)\dag}_{-{1\over2}{1\over2}}  ={2\sqrt{j(j+1)}\over2j+1}\ket{\sfr12,\sfr12}\bra{\sfr12,-\sfr12}\, .
\end{align}
}
The attainability of this bound allows us to express the maximum fidelity as  $F_{\rm max} = 1/3  + {\rm maxeigv} ({\mathsf M})  /6$,  
where ${\rm maxeigv} ({\mathsf M})$ is the maximum eigenvalue of the  $(L+1) \times (L+1)$ tridiagonal symmetric matrix  with entries 
\begin{align}
&\displaystyle {\mathsf M}_{ll}={4l+3\over2l+2},\quad
{\mathsf M}_{ll-1}=\sqrt{{2l-1\over2l}},\quad l\ge2,&\nonumber\\
&\displaystyle {\mathsf M}_{11}={7\over4},\quad
{\mathsf M}_{10}={1\over\sqrt2},\quad
{\mathsf M}_{00}=1.
&
\label{M0 entries}
\end{align}

For small $L$, this maximum eigenvalue can be computed numerically.  For  large $L$, we convert the optimization problem into the extremization of an action functional (see the Supplemental Material \cite{supplemental}), which yields  the closed-form expression (\ref{Fopt}).

{\em The optimal measure-and-operate strategy.}  The optimal quantum strategy  derived in the previous section uses a quantum memory for single atom states.     But what if no quantum memory is available?  
  Is it possible to achieve the same performance by measuring the atom and storing the outcome in a classical memory?     We now show that the answer is negative.
  
  In the lack of quantum memories, the effective channel transforming the sensor's state into an action has the ``measure-and-operate''  form 	$\map{C}_{\rm MO}(\rho)=\sum_{k}  \<  \phi_g|  P_{k} |\phi_g\> ~\map{C}_{k}(\rho)$, where  $(P_k)$ is a positive operator-valued measure (POVM),  representing a quantum measurement on the atom, and the map $\map C_k$ is a quantum channel depending on the measurement outcome and acting on the target system. 
	The fidelity of this strategy can be written as 
	\begin{align}\label{MOfid}
			F^{\rm MO}  (g) \, =\frac 13  +  \, \dfrac{1}{6} \sum_{k}     \<  \phi_g|  P_{k} |\phi_g\>  \,  \bb U_{g}|C_{k}|U_g\kk \,,  
	\end{align}
	where $C_k$ is the Choi operator of the channel $\map C_k$.  
	
The performance of the best measure-and-operate strategy is given by the maximum of Eq.~(\ref{MOfid}) over  $|\phi\>$, $(P_k)$, and $(C_k)$, in the worst case over all rotations $g$.    A crucial simplification comes from the fact that every qubit rotation $U_g$ can be written as $  U_g   = \cos(\theta/2)  \, \id  -  i \sin(\theta/2)  \,  (r_x  X  +  r_y\,  Y +  r_z  Z)$, where $X,Y,Z$ are the three Pauli matrices and $(r_x,r_y,r_z)$	is a vector in $\R^3$.   As a consequence, the Choi operators $C_k$ that maximize  Eq.  (\ref{MOfid}) can be taken without loss of generality to be matrices with real coefficients in the basis $\{|\id\,\kk,    i|  X\kk,  i  |Y\kk ,  i|Z \kk\}$. 	 In turn, this condition implies that the quantum channels $\map C_k$ are unital, that is,  $\map C_k    (\id)=   \id$ for every $k$ (see e.g. Proposition 2 of Ref. \cite{mo2019quantum}).     Finally, we use the fact that all qubit unital channels are random unitary \cite{landau1993birkhoff}. Since the fidelity (\ref{MOfid}) is linear in $\map C_k$, we obtain that the optimal quantum channel $\map C_k$ can be taken to be unitary without loss of generality.  
	
In the Supplemental Material \cite{supplemental}, we show that the optimal measure-and-operate strategy  consists in estimating the unknown rotation $g$  and applying the estimated rotation to the target spin.    	The fidelity for the optimal strategy satisfies the achievable upper bound
\begin{equation}
F_{\max}^{\rm MO}  \le \frac 12  +  \frac 1 6  {\rm maxeigv}  ( {\sf N} )  \, ,   
\label{FMO max}
\end{equation}
where $  {\sf  N}$  is the $(L+1) \times  (L+1)$ tridiagonal matrix with entries  ${\sf N}_{ll}=     l/(l+1)$ and ${\sf N}_{l-1\,l}=  {\sf N}_{l\,l-1}  =  \sqrt{  (2l-1)/(2l+1)}$.   This matrix has a similar form to the matrix $\sf M$ appearing  in the optimization of the quantum strategy.    Using the same techniques, we obtain the asymptotic expression  (\ref{FMO}). Comparing  Eqs.~(\ref{Fopt}) and~(\ref{FMO}) we conclude that the error of the optimal measure-and-operate strategy is twice the error of the optimal quantum strategy.


{\em Conclusions.}  In this paper we identified the ultimate quantum limits for the task of rotating a single  spin based on information gathered by a single atom.  When the atom is initially unentangled with the machine's memory,   a transfer of quantum information from the sensor to the actuators is necessary to achieve the optimal learning performance.   In stark contrast, no transfer of quantum information is required if the initial state of the sensor is entangled with a sufficiently large part of the machine's memory.   Our results demonstrate that the physical properties of the sensor, such as the form of its  coupling with the relevant parameters and its entanglement with other parts of the learning machine can significantly affect the optimal learning strategy.    Overall, this work  opens up an exploration of the interplay between  quantum sensing and quantum learning agents, and of the tradeoff between entanglement and quantum memory as resources for quantum-enhanced learning.


\acknowledgments
 This work was supported by the Chinese Ministry of Science and Technology (MOST) through grant 2023ZD0300600, and by the Hong Kong Research Grant Council (RGC) through grants  SRFS2021-7S02  and R7035-21F, and the State Key Laboratory of Quantum Information Technologies and Materials.   EB acknowledges the financial support of the QuantERA grant C'MON-QSENS!, by Spanish MICINN PCI2019-111869-2 and  the Spanish  Agencia Estatal de Investigaci\'on, project PID2019-107609GB-I00/AEI /10.13039/501100011033.
EB also thanks the hospitality of Computer Science Department of the University of Hong Kong during his stay. 

\nocite{gendra2013optimal}
\nocite{bagan2004dense}

\bibliographystyle{apsrev4-1}


%

  \end{document}


\title{SUPPLEMENTAL MATERIAL FOR \\ Quantum learning with a single-atom sensor}

\author{Mo Yin$^{1}$, Emilio Bagan$^{2,3}$ and Giulio Chiribella$^{3,4,5}$}

\affiliation{
$^1$  Quantum AI Research Lab, Thrust of Artificial Intelligence, Information Hub, The Hong Kong University of Science and Technology (Guangzhou).\\
$^{2}$F\'{i}sica Te\`{o}rica: Informaci\'{o} i Fen\`{o}mens Qu\`antics, Departament de F\'{\i}sica, Universitat Aut\`{o}noma de Barcelona, 08193 Bellaterra (Barcelona), Spain\\
$^{3}$ QICI Quantum Information and Computation Initiative, School of Computing and Data Science,  The University of Hong Kong, Pokfulam Road, Hong Kong\\
$^4$ Quantum Group, Department of Computer Science, University of Oxford, Wolfson Building, Parks Road, Oxford, OX1 3QD, United Kingdom\\
$^5$ Perimeter Institute for Theoretical Physics, 31 Caroline Street North, Waterloo, Ontario, Canada}

\begin{abstract}

This supplemental material provides detailed derivations of the main results regarding the ultimate learning limits of a machine equipped with a single-atom sensor discussed in the main text. It consists of three sections, each devoted to one of the three scenarios mentioned there: the single atom sensor without entanglement with memory, the measure-and-operate strategy, and the sensor entangled~with memory scenario.

Throughout all three sections, we adhere to the same notations and conventions introduced in the main text. To ensure brevity, we avoid reintroducing them in this supplemental material.

\end{abstract}

\maketitle


\section{Optimal strategy with a single atom (no entanglement)}\label{app:optimality}

In Section~\ref{apps:optimal} we provide a precise characterization of the channel $\cal C$ which is then used 
in Section~\ref{apps:max} to derive the upper bound on the fidelity given in Eq.~(\ref{main-optimalF}) of the main text. The asymptotic expression of $F_{\rm max}$ presented in Eq.~(\ref{main-Fopt}) of the main text is derived in Section~\ref{apps:asymp}. The proofs of the various theorems introduced in these sections are collected in Section~\ref{apps:proofs}.

\subsection{\boldmath Characterizing the channel $\cal C$}\label{apps:optimal}

Our starting point is Eq.~(\ref{main-Fe tilde C}) from the main text, which we express as follows:
\begin{equation}
F(g)={1\over3}+{2\over3}\FE(g).
\label{Fe  F relat}
\end{equation}
Here, $\FE(g)$ is the
entanglement fidelity~\cite{horodecki1999general},  given by
\begin{equation}
\FE(g)={1\over4} \bra{\tilde\phi_g}\!\otimes\!\bra{\!\langle U_g} \tilde C \ket{\tilde\phi_g}\!\otimes\!\ket{U_g\rangle\!}.
\label{Fe g}
\end{equation}
Optimizing the fidelity is therefore equivalent to optimizing the entanglement fidelity. Furthermore, the optimization of the worst-case entanglement fidelity, defined as $\FE=\inf_g \FE(g)$, can be achieved by finding the maximum value of
\begin{equation}
\FE={1\over4} \bra{\tilde\phi}\!\otimes\!\bra{\!\langle \id\,} \tilde C \ket{\tilde\phi}\!\otimes\!\ket{\id\,\rangle\!},
\label{Fe wc}
\end{equation}
over all possible Choi operators $C$ and sensor states~$\ket{\phi}$ of the form
\begin{equation}
|\phi\rangle=\sum_{l=0}^L c_j|\phi_l\rangle,\quad \sum_{l=0}^L|c_l|^2=1,
\label{c's}
\end{equation}
as described in Eq.~(\ref{main-c's}) of the main text.
To ease the notation, we drop the subscripts $\rm S$, $T_{\rm in}$ and $T_{\rm out}$ that specify to which Hilbert space the states belong unless confusion may arise (e.g.,  we write $|\id\,\kk$ instead of $|\id\,\kk_{T_{\rm out}T_{\rm in}}$).
In Eq.~(\ref{Fe wc}) the operator $\tilde C$ is chosen to commute with $V_g\otimes U_g\otimes I$ without loss of generality. By Schur's lemma it takes the block diagonal form shown in Eq.~(\ref{main-tilde C block}) of the main text:
\begin{equation}
\tilde C=\bigoplus_{j=1/2}^{J}\id^{(j)}\!\otimes\! B^{(j)},\quad J:=L+\mbox{${1\over2}$} .
\label{tilde C}
\end{equation}
We recall that $B^{(j)}$ acts on the space ${\mathbb C}^{\nu_j}\!\otimes {\mathscr H}_{\rm T_{in}}$, where~$\nu_j$ is the multiplicity of the spin-$j$ representation of the rotation group. In our case, we have $\nu_{J}=1$, and $\nu_j=2$ for \mbox{$j<J$}. So, $B^{(J)}$ is a  $2\times2$ matrix, whereas $B^{(j)}$ are~$4\times4$ matrices, for $j<J$. These operators can be written as
\begin{equation}
B^{(j)}=\sum_{\sigma\sigma'}\ket{j,\sigma}{}\bra{j,\sigma'}\!\otimes\!B^{(j)}_{\sigma\sigma'},
\label{B sigma}
\end{equation}
with
\begin{equation}
B^{(j)}_{\sigma\sigma'}=\sum_{s,s'}B^{(j)}_{\sigma\sigma';ss'}\ket{\mbox{$1\over2$},s}\bra{\mbox{$1\over2$},s'}.
\label{B s s'}
\end{equation}
In Eq.~(\ref{B sigma}), the states $\ket{j,\sigma}$, where $\sigma=\pm\mbox{$1/2$}$ (\mbox{$\sigma=+\mbox{$1/2$}$} for $j=J$), form a basis of the multiplicity space ${\mathbb C}^{\nu_j}$. By convention, the label $\sigma$ indicates that $j$ has been obtained by coupling a spin $\sfr12$ to a~state with orbital angular momentum $l=j-\sigma$. 
With this convention, 
we note that the only possible non-zero element of $B^{(J)}$ is $B^{(J)}_{{1\over2}{1\over2}}$.

In order for $C$ to be a Choi operator of a deterministic channel, it must satisfy the condition
\begin{equation}
\Tr_{\rm T_{out}}[C]=\id_{\rm S}\!\otimes\!\id_{\rm T_{in}}
\label{tr_O},
\end{equation}
which can be expressed as
\begin{align}
&B^{({1\over2})}_{{1\over2}{1\over2}}={1\over2}\id_{\rm T_{in}},\nonumber\\[.5em]
&{2j+3\over2j+2}B^{(j+1)}_{{1\over2}{1\over2}}+{2j+1\over2j+2}B^{(j)}_{-{1\over2}-{1\over2}}=\id_{\rm T_{in}},\quad j<J.
\label{normal}
\end{align}
Since $C$ (and thus $\tilde C$) must be positive semidefinite, additional conditions arise:
\begin{equation}
B^{(J)}_{{1\over2}{1\over2}}\ge0,\quad
\begin{pmatrix}
B^{(j)}_{-{1\over2}-{1\over2}}&B^{(j)}_{-{1\over2}{1\over2}}\\[.7em]
B^{(j)}_{{1\over2}-{1\over2}}&B^{(j)}_{{1\over2}{1\over2}}
\end{pmatrix}
\ge0.
\label{posit}
\end{equation}
In particular, \raisebox{0pt}[0pt][0pt]{$B^{(j)}_{\sigma\sigma}$} must have a spectral representation of the form
\begin{equation}
B^{(j)}_{\sigma\sigma}=\sum_{x=1}^2\lambda^{(j)}_{\sigma x}\;\ketbrad{\beta^{(j)}_{\sigma x}} ,
\label{spectral}
\end{equation}
where $\ket{\beta^{(j)}_{\sigma x}}$, $x=1,2$, are normalized eigenvectors and  $\lambda^{(j)}_{\sigma x}\ge0$. We can pick 
$\lambda^{(j)}_{\sigma1}\ge \lambda^{(j)}_{\sigma2}$ in full generality.
%

After fully parametrizing the set of possible Choi operators $C$, we can now move on to optimizing them. It is tempting to assume that the optimal $B^{(j)}_{\sigma\sigma}$ operators are rank-1 for all $j$ and $\sigma$, except for $\Bpp{{1\over2}}$. In the following subsection we will show that, for the particular problem under consideration, this assumption can be made without loss of generality.

\subsection{Upper bounding the fidelity}\label{apps:max}

Our objective is to derive the upper bound for the (worst case) entanglement fidelity $\FE$ defined in Eq.~(\ref{Fe wc}). 
The derivation is carried out in two steps, referred to as~Theorems~\ref{theor 1} and~\ref{theor 2} below. The complete proofs of these theorems are given in Section~\ref{apps:proofs}. The attainability of the bound is discussed  in the main text.


The bound
provided by our first theorem is expressed in terms of the sensor state components $c_l$, which can be assumed to be non-negative, and holds true for generic normalized states $\{\ket{\phi_l}\}_{l=0}^L$ [refer to Eq.~(\ref{main-c's}) of the main text]. The bound also depends on the choice of $C$ through the eigenvalue \raisebox{0ex}[2.3ex][1.1ex]{$\lambda^{({3\over2})}_{{1\over2}1}$} and the smallest eigenvalue of each $\Blpp{j}$ [note that the eigenvalues of \raisebox{0ex}[0ex][0ex]{$B^{(j)}_{\sigma\sigma}$} are not independent due to the normalization condition in Eq.~(\ref{normal})]. To simplify our expressions, we introduce lighter notation in Theorem~\ref{theor 1}:
\begin{equation}
\lambda_{l}:={2l+2\over2l+1}\lambda^{(l+{1\over2})}_{{1\over2}2}, \qquad \lambda'_{1}:={4\over3}\lambda^{({3\over2})}_{{1\over2}1}.
\label{renorm lambdas}
\end{equation}
These parameters are normalized such that $0\le\lambda_l\le1$, $l=1,2,\dots, L$, and $0\le\lambda'_1\le1$. The parameter $\lambda'_1$ arises because $l=0,1$ are special. As indicated in Eq.~(\ref{normal}), the normalization condition  fixes \raisebox{0ex}[0ex][1ex]{$\Bpp{{1\over2}}$} to be~$\id_{\rm T_{in}}/2$, and this operator is coupled to the components $l=0,1$ of $\ket\phi$. %

\begin{theorem}\label{theor 1}
For any given set of components~$\{c_l\}_{l=0}^L$ and normalized states~$\{\ket{\phi_l}\}_{l=0}^L$ in~(\ref{c's}), and any choice of Choi operator $C$, the corresponding entanglement fidelity~$\FE$ in Eq.~(\ref{Fe wc}) can be upper bounded~as follows
\begin{multline}
\FE\le {1\over4}\sum_{l=2}^L c_l^2\left({4l+3\over2l+2}-{2l+1\over2l+2}\lambda_l\right)\\
+{1\over2}\sum_{l=2}^Lc_lc_{l-1}\left(\sqrt{{2l-1\over2l}}+{\lambda_{l-1}\over2\sqrt{2l(2l-1)}}\right)\\
+{1\over4}c_1^2\left(1+{3\over4} \lambda'_1-{3\over4}\lambda_1\right)\\
+{1\over2}c_1c_0\sqrt{1-{1\over2}\lambda'_1-{1\over2}\lambda_1}+{1\over4}c_0^2 ,
\label{B theor 1}
\end{multline}
where the dependence on $C$ is through $\lambda'_1$ and $\lambda_l$, defined in Eq.~(\ref{renorm lambdas}). 
\end{theorem}
%


Our next theorem addresses the optimization of both the components $c_l$ of the sensor state and the choice of~the Choi operator $C$. It provides an upper bound on the~maximum $\FE$, which is given by the expression on the right-hand side of Equation (\ref{B theor 1}) with $\lambda'_1=1$ and $\lambda_l=0$ for $l=1,2,\dots, L$. 
In essence, this result confirms the validity of the assumption that rank-1 operators $B^{(j)}_{\sigma\sigma}$ are optimal for $j>1/2$.

\begin{theorem}\label{theor 2}
If $\{c_l\}_{l=0}^L$ are the components of the optimal sensor state~$|\phi\rangle$, the maximum value of the expression on the right hand side of~(\ref{B theor 1}) in~Theorem~\ref{theor 1} is attained when $\lambda'_1=1$ and $\lambda_l=0$ for all $l=1,2,\dots,L$. As a result, we obtain an upper bound for the entanglement fidelity $\FE$ that reads:
\begin{multline}
\FE \le\max_{\{c_l\}}{1\over4}\left(\sum_{l=2}^L c_l^2{4l+3\over2l+2}\right.\\
\left.+2\sum_{l=2}^L\!c_lc_{l-1}\sqrt{{2l\!-\!1\over2l}}
\!+\! {7\over4} c_1^2
\!+\!\sqrt2 c_1c_0\!+\!c_0^2\!
\right).
\end{multline}
\end{theorem}

 Eq.~(\ref{main-optimalF}) in the main text directly follows from Theorem~\ref{theor 2}, taking into account Eqs. (\ref{Fe F relat}) and (\ref{Fe wc}).

\subsection{Asymptotic fidelity}\label{apps:asymp}

Recalling the attainability of the upper bound given in Eq.~(\ref{main-optimalF}) of the main text, the maximum fidelity can be expressed~as
\begin{equation}
F_{\rm max} ={1\over 3}+{1\over6}\operatorname{maxeigv}{\mathsf M},
\label{F & maxeigv}
\end{equation}
where ${\mathsf M}$ is the $(L+1)\times(L+1)$ tridiagonal symmetric matrix. Its non-zero independent entries are given by
\begin{align}
&\displaystyle {\mathsf M}_{ll}={4l+3\over2l+2},\quad
{\mathsf M}_{ll-1}=\sqrt{{2l-1\over2l}},\quad l\ge2,&\nonumber\\
&\displaystyle {\mathsf M}_{11}={7\over4},\quad
{\mathsf M}_{10}={1\over\sqrt2},\quad
{\mathsf M}_{00}=1.
&
\label{M0 entries}
\end{align}

Our objective is to derive the asymptotic expression of $\mu:=\operatorname{maxeigv}({\sf M})$ for large $L$. To accomplish this, we rely on the key result provided by the following theorem, the proof of  which is presented in the next section.
\begin{theorem}\label{theor 5}
The asymptotic behavior of the largest eigenvalue of the matrix $\mathsf M$, defined in Eq.~(\ref{M0 entries}), is given~by
\begin{equation}
\mu\sim 4-{1\over L}+{\gamma_1\over L^{4/3}}-{{8\over15}\gamma_1^2\over L^{5/3}}+O(L^{-2}),
\label{expr theor 5}
\end{equation}
where $\gamma_1\approx -2.3381$ is the first zero of the Airy function of the first kind, $\operatorname{Ai}(x)$.
\end{theorem}
\begin{figure}[htbp]
\begin{center}
\includegraphics[scale=.6]{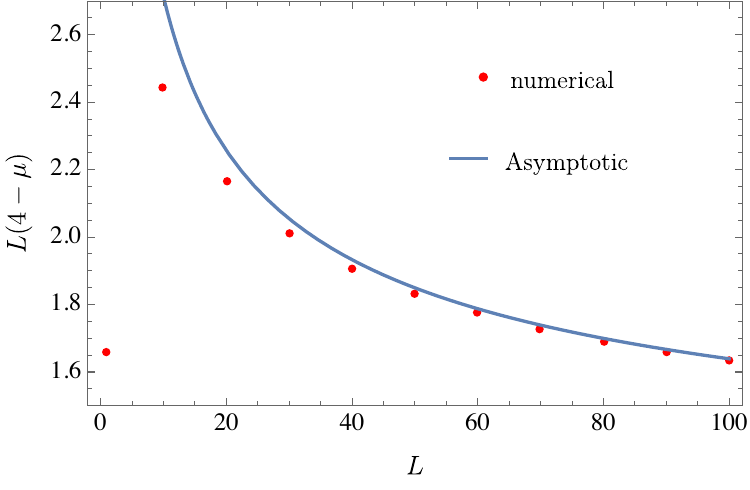}
\caption{Plot of the function $L(4-\mu)$ vs $L$ for $L$ up to~$100$. The solid line is derived from the asymptotic expression in Eq.~(\ref{expr theor 5}), i.e.,  $L(4-\mu)=1-\gamma_1 L^{-1/3}+(8/15)\gamma_1^2 L^{-2/3}$, whereas the red circles are obtained by numerical calculation. The plot illustrates the validity of Eq.~(\ref{expr theor 5}) even for relatively small values of $L$.\label{fig1}}
\label{default}
\end{center}
\end{figure}

Fig~\ref{fig1} illustrates the accuracy  of the asymptotic expression provided by Theorem~\ref{theor 5} 
by comparing the subleading terms in Eq.~(\ref{expr theor 5}) to the exact values of $L(4-\mu)$ obtained through numerical calculation of the largest eigenvalue. The agreement between the asymptotic formula and the numerical results is remarkable. The asymptotic expression accurately predicting $L(4-\mu)$ for values of the total angular momentum as small as $L\gtrapprox 10$, despite of the fact that this comparison accentuates the differences between the two curves.

Applying Theorem~\ref{theor 5}, we can directly conclude that, up to order $1/L^2$, the asymptotic fidelity is given by Eq.~(\ref{main-Fopt}) in the main text. This result provides a concise and accurate expression for estimating the fidelity in the large~$L$ limit.

\subsection{Proof of Theorems~\ref{theor 1}--\ref{theor 5}}\label{apps:proofs}

\subsubsection{Proof of Theorem~\ref{theor 1}}
We start by  inserting
\begin{align}
\id&=\bigoplus_{j=1/2}^J\sum_\sigma\id^{(j)}\otimes\ketbrad{j,\sigma}\otimes\id_{\rm T_{in}}\nonumber\\
&:=
\bigoplus_{j=1/2}^J\sum_{\sigma}\id^{(j)}_{\sigma}
\end{align}
on both sides of $\tilde C$ in Eq.~(\ref{Fe wc}). We can then use the Schwarz inequality in the obvious way on the terms with $j\ge3/2$, to upper bound the fidelity as
\begin{multline}
\FE_L\!\le\!{1\over4}\Bigg\{\sum_{\sigma,\sigma'}\!\left|\bra{\tilde\phi}\!\otimes\!\bra{\!\<\id\,}
\id^{({1\over2})}_{\sigma}\!\tilde C\id^{({1\over2})}_{\sigma'}
\ket{\tilde\phi}\!\otimes\!\ket{\id\,\>\!}\right|\!\\ 
+\sum_{j=3/2}^J\!\!\left( w^{(j)}_{1\over2}\!+w^{(j)}_{-{1\over2}}\right)^2
\!\!\Bigg\}
\label{schw},
\end{multline}
where
\begin{equation}
w^{(j)}_{\sigma}:=\left\|\sqrt{\tilde C}\, \id^{(j)}_{\sigma}\ket{\tilde\phi}\!\otimes\!\ket{\id\,\>\!}\right\|.
\end{equation}
We collect the various terms according to
\begin{align}
\FE_L&\!\le\!{1\over4}\Bigg\{\sum_{\sigma,\sigma'}\!\left|\bra{\tilde\phi}\!\otimes\!\bra{\!\<\id\,}
\id^{({1\over2})}_{\sigma}\!\tilde C\id^{({1\over2})}_{\sigma'}
\ket{\tilde\phi}\!\otimes\!\ket{\id\,\>\!}\right|\!+\!\left[\!w^{({3\over2})}_{1\over2}\!\right]^2\!\!\Bigg\}\nonumber\\
&+\!{1\over4}\!\left\{\!\sum_{j=5/2}^J\!\! \left[\!w^{(j)}_{1\over2}\!\right]^2\!\!+\!\!\!\sum_{j=3/2}^J\!\!\!\left(\!2w^{(j)}_{{1\over2}}w^{(j)}_{-{1\over2}}
\!+\!\left[\!w^{(j)}_{-{1\over2}}\!\right]^2\right)
\!\!\right\}
\nonumber\\
&:=\fe+\Fe
\label{F1/2 F>1/2},
\end{align}
In order to obtain an attainable bound, the terms in the first line of Eq.~(\ref{F1/2 F>1/2}), which we denote by $\fe$, need special consideration. 
So, we will first focus on upper bounding the terms in the second line, denoted by  $\Fe$, for which the procedure is less involved.

Using the spectral representation in Eq.~(\ref{spectral}), we readily see that
\begin{equation}
\left[w^{(j)}_\sigma\right]^2=\sum_{x=1}^2 \lambda^{(j)}_{\sigma x} \bra{\tilde\phi}\otimes\bra{\bar\beta^{(j)}_{\sigma x}}\,\id^{(j)}_{\sigma}
\ket{\tilde\phi}\otimes\ket{\bar\beta^{(j)}_{\sigma x}}.
\label{the w's}
\end{equation}
In turn, we can bound these norms as
\begin{align}
w^{(j)}_{{1\over2}}&\le |c_{j-{1\over2}}|\sqrt{\lambda^{(j)}_{{1\over2}1}+{\lambda^{(j)}_{{1\over2}2}\over2j}},
\label{W+}\\[.5em]
w^{(j)}_{-{1\over2}}&\le |c_{j+{1\over2}}|\sqrt{{2j+1\over2j+2}\lambda^{(j)}_{-{1\over2}1}},
\label{W-}
\end{align}
where we have  used the next lemma.

\begin{lemma}\label{lemma1}
Let $\ket{\beta_x}$, $x=1,2$, be two orthogonal qubit states and~$p_x$, $x=1,2$ two non-negative numbers such that $p_1\ge p_2$ and $p_1+p_2=1$. Let $\ket{\phi_l}$ be a normalized state in the spin $l$ representation of {\em SU(2)}. Then, we have the following two attainable bounds:
{\em
\begin{align}
\sum_xp_x\bra{\phi_{j-{1\over2}}}\!\otimes\!\bra{\beta_x}\id^{(j)}_{{1\over2}}\ket{\phi_{j-{1\over2}}}\!\otimes\!\ket{\beta_x}&\le
p_1+{p_2\over 2l},\label{lemma 1 1}\\[.05em]
\sum_xp_x\bra{\phi_{j+{1\over2}}}\!\otimes\!\bra{\beta_x}\id^{(j)}_{-{1\over2}}\ket{\phi_{j+{1\over2}}}\!\otimes\!\ket{\beta_x}&\le
{2j\!+\!1\over2j\!+\!2}p_1.\label{lemma 1 2}
\end{align}
}
\end{lemma}

\noindent{\bf Proof.} Since the quantities we wish to evaluate are invariant under rotations, we can choose without loss of generality
$\ket{\beta_1}=\ket{\sfr12 ,\sfr12 }$ and $\ket{\beta_2}=\ket{\sfr12 ,-\sfr12 }$. Let us denote the left hand side of Eq.~(\ref{lemma 1 1}) by $v_{{1\over2}}$. Then, using the Clebsch-Gordan coefficients we have
\begin{align}
v_{{1\over2}}&=\sum_{m}\left | a_{m}\right|^2\sum_x
p_x{2j-2(-1)^xm+1\over4j}\nonumber\\
&\le \max_{m}\;
\sum_x
p_x{2j-2(-1)^xm+1\over4j}\nonumber\\
&=p_1+{p_2\over2j},
\end{align}
where the range of $m$ is $-j+\sfr12\le m\le j-\sfr12$ and $a_m$ are the components of the normalized state $\ket{\phi_{j-{1\over2}}}$.
Similarly, let us denote the left hand side of Eq.~({\ref{lemma 1 2}}) by $v_{-{1\over2}}$. One has
\begin{align}
v_{-{1\over2}}&=\sum_{m}\left | a_{m}\right|^2\sum_x
p_x{2j+2(-1)^xm+1\over4(j+1)}\nonumber\\
&\le \max_{m}\;
\sum_x
p_x{2j+2(-1)^xm+1\over4j}\nonumber\\
&={2j+1\over2j+2}p_1,
\end{align}
where in this case the range of $m$ is $-j-\sfr12\le m\le j+\sfr12$ and $a_m$ are the components of $\ket{\phi_{j+{1\over2}}}$~$\blacksquare$


We next express the normalization condition in Eq.~(\ref{normal}) in terms of the eigenvalues $\lambda^{(j)}_{\sigma x}$. 
This equation implies that the operators $\Blpp{j\!+\!1}$ and $\Blmm{j}$ share the same eigenvectors. In particular, we have the following eigenvalue equation for $B^{(j+1)}_{{1\over2}{1\over2}}$:
\begin{multline}
B^{(j+1)}_{{1\over2}{1\over2}}|\beta^{(j)}_{-{1\over2}x}\rangle={2j+2-(2j+1)\lambda^{(j)}_{-{1\over2}x}\over 2j+3}|\beta^{(j)}_{-{1\over2}x}\rangle .
\label{share eigenvec}
\end{multline}
Our labeling convention, $\lambda^{(j)}_{\sigma1}\ge\lambda^{(j)}_{\sigma2}$ [first paragraph after Eq.~(\ref{spectral})], requires that
\begin{equation}
|\beta^{(j+1)}_{{1\over2}x}\rangle={\rm e}^{i\alpha_x}|\beta^{(j)}_{-{1\over2}\bar x}\rangle,\quad
\lambda^{(j+1)}_{{1\over2}x}=\lambda^{(j)}_{-{1\over2}\bar x},
\label{eigenvec eigenval}
\end{equation}
where $0\le\alpha_x<2\pi$ and $\bar x=3-x$ [i.e., $\bar x={\sf NOT}(x)$]. Hereafter we set $\Llm{j}{2}=0$, for $j>\sfr12$, since these eigenvalues do not appear in our bound to the fidelity [see Eqs.~(\ref{W+}) and~(\ref{W-})]. With this condition it also follows from Eq.~(\ref{share eigenvec}) that
\begin{equation}
\lambda^{(j)}_{{1\over2}1}={2j\over 2j+1},\quad
\lambda^{(j)}_{-{1\over2}1}={2j+2\over 2j+1}-{2j+3\over 2j+1}\lambda^{(j+1)}_{{1\over2}2},
\label{normal lambda}
\end{equation}
for  $j=\sfr12,\sfr32,\dots,J-1$ and $\lambda^{({1\over2})}_{{1\over2}2}=1/2$. Using these last equations, we can conveniently upper bound~$\Fe$ in Eq.~(\ref{F1/2 F>1/2}) in terms of the coefficients $c_l$ and the eigenvalues $\Llp{j}{2}$ alone. In particular, 
choosing these eigenvalues to be zero would imply that the operators $\Blpp{j\!+\!1}$ and $\Blmm{j}$ are rank-1. Unfortunately, at this stage it is not clear whether this is the optimal choice and we need to keep these eigenvalues as free parameters throughout the entire optimization. 

With the above, we have
\begin{align}
\Fe\!&\le\!
{1\over2}\!\!\sum_{j=3/2}^{J-1}\!\!c_{j-{1\over2}}c_{j+{1\over2}}\sqrt{\!\left(\!\!1\!-\!{2j\!+\!3\over2j\!+\!2}\lambda^{(j+1)}_{{1\over2}2}\!\right)\!\!\!
\left(\!{2j\over2j\!+\!1}\!+\!{\lambda^{(j)}_{{1\over2}2}\over2l}\!\right)\!}\nonumber\\
&+\!{1\over4}\!\!\sum_{j=5/2}^J\!\!c^2_{j-{1\over2}}\left({4j+1\over2j+1}-\lambda^{(j)}_{{1\over2}2}\right),
\label{B1J bound}
\end{align}
where we have chosen $c_l$ to be positive with no loss of generality. From the normalization conditions, Eq.~(\ref{normal lambda}), we have $0\le\lambda^{(j)}_{{1\over2}2}\le(2j)/(2j+1)$.

We now consider $\fe$, whose expression is given in Eq.~(\ref{F1/2 F>1/2}) (top line). We can also write
\begin{multline}
\fe\!=\!{1\over4}\Bigg\{\!
\left[\!w^{({1\over2})}_{1\over2}\!\right]^2\!\!+\left[\!w^{({1\over2})}_{-{1\over2}}\!\right]^2\!\!+\left[\!w^{({3\over2})}_{1\over2}\!\right]^2
\\
+
2\!\left|\bra{\tilde\phi}\!\otimes\!\bra{\!\<\id\,}
\id^{({1\over2})}_{{1\over2}}\tilde C\id^{({1\over2})}_{-{1\over2}}
\ket{\tilde\phi}\!\otimes\!\ket{\id\,\>\!}\right|\!
\!\Bigg\}
\label{B01}.
\end{multline}
The first  term is easily computed, as we recall that the normalization condition implies $\Bpp{{1\over2}}=\id_{\rm T_{in}}/2$ [equivalently, one can check that Eq.~(\ref{normal lambda}) gives $\Lp{{1\over2}}{x}=1/2$] and $\ket{\phi_0}=\ket{j=0,m=0}$. We obtain
\begin{equation}
\left[\!w^{({1\over2})}_{1\over2}\!\right]^2=c_0^2.
\end{equation}

Using the next lemma, we can bound the other two terms in the first line of Eq.~(\ref{B01}).
\begin{lemma}\label{lemma 2}
With the definitions of Eqs.~(\ref{spectral}) and~(\ref{the w's}) one has the upper bound
\begin{equation}
\left[w^{({1\over2})}_{-{1\over2}}\right]^2\!\!+\!\left[w^{({3\over2})}_{1\over2}\right]^2\le c_1^2\left(
1\!+\!\lambda^{({3\over2})}_{{1\over2}1}\!-\!\lambda^{({3\over2})}_{{1\over2}2}\right).
\label{bound lemma 2}
\end{equation}
\end{lemma}

\noindent{\bf Proof.} Rather than using the bounds provided by Lemma~\ref{lemma1}, we compute the two terms on the left hand side explicitly. Rotational invariance allows us to choose the eigenvectors of $\Bpp{{3\over2}}$ and $\Bmm{{1\over2}}$ to be
\raisebox{0pt}[0pt][5pt]{$\ket{\sfr12 ,\pm\sfr12}$},  as we did to prove Lemma~\ref{lemma1}. More precisely, we choose 
$|\beta^{({3\over2})}_{{1\over2}1}\rangle=|\beta^{({1\over2})}_{-{1\over2}2}\rangle=|{1\over2},{1\over2}\rangle$ and $|\beta^{({3\over2})}_{{1\over2}2}\rangle=|\beta^{({1\over2})}_{-{1\over2}1}\rangle=|{1\over2},-{1\over2}\rangle$.
Using also the required Clebsch-Gordan coefficients, we obtain
\begin{multline}
\left[w^{({3\over2})}_{1\over2}\right]^2=
c_1^2 \left[ \lambda^{({3\over2})}_{{1\over2}1}\left(| a_{1}|^2+\sfr23 | a_{0}|^2+\sfr13 | a_{-1}|^2 \right)\right.\\
+\left.
 \lambda^{({3\over2})}_{{1\over2}2}\left(\sfr13 | a_{1}|^2+\sfr23 | a_{0}|^2+ | a_{-1}|^2 \right)\right],
 \label{w3/2}
\end{multline}
where $a_m$, $m=-1,0,1$, are the components of the normalized state $\ket{\tilde \phi_1}$. 
Likewise, 
\begin{multline}
\left[w^{({1\over2})}_{-{1\over2}}\right]^2=c_1^2 \left[ \lambda^{({1\over2})}_{-{1\over2}1}\left(\sfr13 | a_{0}|^2+\sfr23 | a_{1}|^2 \right)\right.\\
+\left.
 \lambda^{({1\over2})}_{-{1\over2}2}\left(\sfr13 | a_{0}|^2+\sfr23  | a_{-1}|^2 \right)\right].
 \label{diag 1}
\end{multline}
%
Combining Eqs.~(\ref{w3/2}) and~(\ref{diag 1}) we obtain
\begin{multline}
\left[w^{({3\over2})}_{1\over2}\right]^2\!\!+\!\left[w^{({1\over2})}_{-{1\over2}}\right]^2
\!\!=\\
c_1^2\!\left[1\!+\!\left(\lambda^{({3\over2})}_{{1\over2}1}\!-\!\lambda^{({3\over2})}_{{1\over2}2}\right)\left(| a_1|^2\!-\!| a_{-1}|^2\right)\right],
\end{multline}
where we have made use of the normalization condition in Eq.~(\ref{normal}) [equivalently, of Eq.~(\ref{share eigenvec}) and Eq.~(\ref{eigenvec eigenval})] to trade the eigenvalues $\Lm{{1\over2}}{1}$ and $\Lm{{1\over2}}{2}$ for  $\Lp{{3\over2}}{1}$ and $\Lp{{3\over2}}{2}$. Since~$\Lp{{3\over2}}{1}\ge\Lp{{3\over2}}{2}$, we obtain the bound in Eq.~(\ref{bound lemma 2})~$\blacksquare$


Finally, we can bound the last term in Eq.~(\ref{B01}) using the following lemma.

\begin{lemma}\label{lemma 3}
With the definitions  in Eqs.~(\ref{B sigma}),~(\ref{B s s'}) and~(\ref{spectral}), and recalling Eq.~(\ref{tilde C}), we have the following bound
{\em
\begin{multline}
\left|\bra{\tilde\phi}\!\otimes\!\bra{\!\<\id\,}
\id^{({1\over2})}_{{1\over2}}\tilde C\id^{({1\over2})}_{-{1\over2}}
\ket{\tilde\phi}\!\otimes\!\ket{\id\,\>\!}\right|\\
\le c_0c_1\sqrt{1-\sfr23 \lambda^{({3\over2})}_{{1\over2}1}-\sfr23 \lambda^{({3\over2})}_{{1\over2}2}}.
\label{wpm lemma 3}
\end{multline}
}
\end{lemma}

\noindent{\bf Proof.} 
Let us denote the left hand side of Eq.~(\ref{wpm lemma 3}) by~$W$ and write 
\begin{equation}
B^{({1\over2})}_{{1\over2}-{1\over2}}=\sum_{s,s'=-{1\over2}}^{{1\over2}}\gamma_{ss'}\ket{\sfr12, s}\bra{\sfr12, s'}.
\end{equation}
Sticking to the same notation as in Lemma~\ref{lemma 2} and using the appropriate Clebsch-Gordan coefficients, we have 
\begin{align}
W&={1\over2}c_0c_1\left|
-\sfr{1}{\sqrt3} a_0\gamma_{{1\over2}{1\over2}}
+\sfr{\sqrt2}{\sqrt3} a_{1}\gamma_{-{1\over2}{1\over2}}\right.\nonumber\\
&+\left.
\sfr{1}{\sqrt3} a_0\gamma_{-{1\over2}-{1\over2}}
-\sfr{\sqrt2}{\sqrt3} a_{-1}\gamma_{{1\over2}-{1\over2}}
\right|\nonumber\\
\le\,&{1\over2}c_0c_1\!
\left[\!
\sqrt{\!\left(|\gamma_{{1\over2}{1\over2}}|^2\!+\!|\gamma_{-{1\over2}{1\over2}}|^2\!\right)\!\left(
\sfr{1}{3}| a_0|^2\!+\!
\sfr{2}{3}| a_{1}|^2
\right)}
\right.\nonumber\\
&+\left.
\!\!\sqrt{\!\left(|\gamma_{-{1\over2}-{1\over2}}|^2\!+\!|\gamma_{{1\over2}-{1\over2}}|^2\!\right)\!\left(
\sfr{1}{3} | a_0|^2\!+\!
\sfr{2}{3} | a_{-1}|^2
\right)\!}\,
\right],
\label{off 1}
\end{align}
where we note that we have used Schwarz inequality on both the first and second lines in Eq.~(\ref{off 1}).
This expression can be further simplified by noticing that we can set $ a_0=0$ with no loss of generality (just make the substitution $| a_{\pm1}|^2\!+\!\sfr12 | a_0|^2\!\to\! | a_{\pm1}|^2$). 

We can now get rid of the gamma coefficients by recalling the positivity condition in Eq.~(\ref{posit}) for $j=1/2$. In terms of  the parameters that we have introduced above, it can be written as
\begin{eqnarray}
&\displaystyle 
\begin{pmatrix}
\lambda^{({1\over2})}_{-{1\over2}2}&0&\bar\gamma_{{1\over2}{1\over2}}&\bar \gamma_{-{1\over2}{1\over2}}\\
0&\lambda^{({1\over2})}_{-{1\over2}1}&\bar\gamma_{{1\over2}-{1\over2}}&\bar\gamma_{-{1\over2}-{1\over2}}\\
\gamma_{{1\over2}{1\over2}}&\gamma_{{1\over2}-{1\over2}}&{1\over2}&0\\
\gamma_{-{1\over2}{1\over2}}&\gamma_{-{1\over2}-{1\over2}}&0&{1\over2}
\end{pmatrix} 
\ge0 .\\
&\nonumber
\end{eqnarray}
Two necessary conditions follow by computing the $(1,1)$-minor  and the $(2,2)$-minor of this matrix:
\begin{align}
|\gamma_{-{1\over2}-{1\over2}}|^2+\!|\gamma_{{1\over2}-{1\over2}}|^2&\le\!{\lambda^{({1\over2})}_{-{1\over2}1}\over 2},\nonumber\\
|\gamma_{{1\over2}{1\over2}}|^2+\!|\gamma_{-{1\over2}{1\over2}}|^2&\le\!{\lambda^{({1\over2})}_{-{1\over2}2}\over 2}.
\end{align}
We now use these inequalities in Eq.~(\ref{off 1}) to further upper bound Eq.~(\ref{off 1}) by
\begin{multline}
{1\over2\sqrt3}\,c_0c_1\!\left(\sqrt{\lambda^{({1\over2})}_{-{1\over2}2}}|a_{1}|+\sqrt{\lambda^{({1\over2})}_{-{1\over2}1}}|a_{-1}|\right)\\
\le{1\over2\sqrt3}\,c_0c_1\!\sqrt{ \lambda^{({1\over2})}_{-{1\over2}1}+\lambda^{({1\over2})}_{-{1\over2}2}},
\end{multline}
where we have used again Schwarz inequality and the fact that $\ket{\tilde \phi_1}$ has unit length. Finally, we substitute the lambdas appearing in this equation by those in Eq.~(\ref{wpm lemma 3}) by recalling the normalization condition in Eq.~(\ref{normal}), as in the proof of Lemma~\ref{lemma 2}~$\blacksquare$


Putting all the pieces together we obtain
\begin{multline}
\fe\le {1\over4}c_0^2+{1\over4}c_1^2\left(1\!+\!\lambda^{({3\over2})}_{{1\over2}1}\!-\!\lambda^{({3\over2})}_{{1\over2}2}\right)\\
+{1\over2}c_0c_1 \sqrt{1-\sfr23 \lambda^{({3\over2})}_{{1\over2}1}-\sfr23 \lambda^{({3\over2})}_{{1\over2}2}}.
\end{multline}
Substituting this and Eq.~(\ref{B1J bound}) into Eq.~(\ref{F1/2 F>1/2}) we find the upper bound we are seeking for. To ease a bit the notation, we introduce the new parameter $\lambda_l$ and $\lambda'_1$ in Eq.~(\ref{renorm lambdas}). 
Then, we have
\begin{multline}
\FE\le{1\over4}\sum_{l=2}^L c_l^2\left({4l+3\over2l+2}-{2l+1\over2l+2}\lambda_l\right)\\
+{1\over2}\sum_{l=2}^Lc_lc_{l-1}\sqrt{1-\lambda_l}\sqrt{{2l-1\over2l}+{\lambda_{l-1}\over2l}}\\
+{1\over4}c_1^2\left(1+{3\over4} \lambda'_1-{3\over4}\lambda_1\right)\\
+{1\over2}c_1c_0\sqrt{1-{1\over2}\lambda'_1-{1\over2}\lambda_1}+{1\over4}c_0^2,
\end{multline}
and $0\le\lambda_l\le1$ and $0\le\lambda'_1\le1$. The very last simplification in this lengthly derivation consists in applying the bounds $\sqrt{1-\lambda_l}\le 1$  and $\sqrt{a+b}\le \sqrt a+b/(2\sqrt{a})$, which hold for any $a>0$ and any $b\ge0$, on the second line of this inequality to obtain the desired result in Eq.~(\ref{B theor 1}).

\subsubsection{Proof of Theorem~\ref{theor 2}}

Let us denote the right hand side of~(\ref{B theor 1}) by ${\mathscr B}$. It can be written in matrix form as
\begin{equation}
{\mathscr B}={1\over4}{\mathsf c}^t{\mathsf M}{\mathsf c},
\end{equation}
where, as above, 
${\mathsf c}^t=(c_L,c_{L-1},\dots, c_1,c_0)$ and ${\mathsf M}$ is the tridiagonal symmetric matrix of dimension~$L+1$ whose non-zero independent entries are
\begin{align}
{\mathsf M}_{ll}&={4l+3\over2l+2}-{2l+1\over2l+2}\lambda_l,\nonumber\\
{\mathsf M}_{ll-1}&=\sqrt{{2l-1\over2l}}+{\lambda_{l-1}\over2\sqrt{2l(2l-1)}},\nonumber\\
{\mathsf M}_{11}&=1+{3\over4} \lambda'_1-{3\over4}\lambda_1,\nonumber\\
{\mathsf M}_{10}&=\sqrt{1-{1\over2}\lambda'_1-{1\over2}\lambda_1},\nonumber\\
{\mathsf M}_{00}&=1,
\label{M entries}
\end{align}
and $l\ge2$  in the top two lines.  Then, the optimal ${\mathsf c}$ is the eigenvector of ${\mathsf M}$ corresponding to its largest eigenvalue (the largest eigenvector for short), which we simply call~$\mu$. In turn, $\mu/4$ gives the maximum value of~${\mathscr B}$ when the components~$c_l$ are optimized. 


We will prove Theorem~\ref{theor 2} in two steps, the first of which we present as Lemma~\ref{lemma 4} below.
\begin{lemma}\label{lemma 4}
The maximum value of the largest eigenvalue, $\mu$, of the matrix ${\mathsf M}$ defined in Eq.~(\ref{M entries}) is attained when $\lambda_l=0$ for $2\le l\le L$.
\end{lemma}

\noindent{\bf Proof.} Obviously, the choice $\lambda_L=0$ is optimal, since ${\mathsf M}_{LL}$ is a decreasing function of $\lambda_L$ and the other entries of ${\mathsf M}$ are independent of~$\lambda_L$.  


Since $\mu$ is an increasing function of $L$ and  ${\mathsf M}$ is non-negative (all of its entries are non-negative) we have the upper bound
\begin{equation}
\mu < \lim_{L\to\infty}\max_{l'}\,\sum_{l=0}^L {\mathsf M}_{ll'}=4.
\end{equation}
The eigenvalue equation ${\mathsf M}{\mathsf c}=\mu{\mathsf c}$ implies
\begin{equation}
{\mathsf M}_{ll} c_l+ {\mathsf M}_{ll-1} c_{l-1}+{\mathsf M}_{l+1l} c_{l+1} =\mu c_l
\end{equation}
for $1\le l\le L-1$. Since all the terms are positive, we can drop the second one and write
\begin{equation}
{\mathsf M}_{ll} c_l+ {\mathsf M}_{l+1l} c_{l+1}\le\mu c_l .
\end{equation}
Substituting the explicit expressions of the matrix entries, collected in Eq.~(\ref{M entries}), we obtain
\begin{equation}
c_{l+1}\le ({\mu-1})\sqrt{2l+2\over 2l+1}c_l\le 3\sqrt{2l+2\over 2l+1}c_l.
\label{bound cj+1}
\end{equation}
Let ${\mathsf M}^{(0)}$ be the matrix obtained by setting $\lambda_l=0$, $2\le l\le L$, and $\mu^{(0)}$ its largest eigenvalue. We have
\begin{multline}
\mu={\mathsf c}^t {\mathsf M}^{(0)}{\mathsf c}+{\mathsf c}^t\left({\mathsf M}-{\mathsf M}^{(0)}\right){\mathsf c}
\\
\le\mu^{(0)}+{\mathsf c}^t\left({\mathsf M}-{\mathsf M}^{(0)}\right){\mathsf c},
\label{mu mu0}
\end{multline}
as ${\mathsf c}$ is the largest eigenvector of ${\mathsf M}$, but not necessarily of ${\mathsf M}^{(0)}$. The last term can be easily bounded using Eq.~(\ref{bound cj+1}), and we can write
\begin{equation}
{\mathsf c}^t\left({\mathsf M}-{\mathsf M}^{(0)}\right){\mathsf c}\le\sum_{l=2}^{L-1}
{2l+5-4l^2\over(2l+1)(2l+2)}\lambda_l c_l^2.
\end{equation}
Each term of this sum is less or equal to zero, hence we have $\mu\le\mu^{(0)}$~$\blacksquare$


Now that we know that the maximum of ${\mathscr B}$ is attained by $\lambda_l=0$ for $2\le l\le L$, we turn our attention to the optimal value of 
$\lambda'_1$. 


To avoid introducing further notation, in what follows let ${\mathsf M}$ stand for the original matrix in Eq.~(\ref{M entries}) with $\lambda_L=\lambda_{L-1}=\cdots=\lambda_2=0$.
The eigenvalue equation tells us that the components $c_0$ and $c_1$ of its largest eigenvector must satisfy
\begin{equation}
{\mathsf M}_{10}c_1+{\mathsf M}_{00}c_0=\mu c_0.
\label{link c1 c0}
\end{equation}
Then,
\begin{equation}
c_1={\mu-1\over\sqrt{1-\sfr12 \lambda'_1-\sfr12 \lambda_1}}c_0.
\label{c1 c0}
\end{equation}
Let ${\mathsf M}^{(0)}$ be now the matrix ${\mathsf M}$ for $\lambda_1'=1$ 
and~$\mu^{(0)}$ its corresponding largest eigenvalue. Then, Eq.~(\ref{mu mu0}) also holds for these redefinitions of ${\mathsf M}$, ${\mathsf M}^{(0)}$, $\mu$ and $\mu^{(0)}$.  
Using Eq.~(\ref{c1 c0}) we note that ${\mathsf c}^t({\mathsf M}-{\mathsf M}^{(0)}){\mathsf c}\le0$ is now equivalent to the condition
\begin{multline}
(3\mu-7)\lambda'_1-3\mu+11-4\lambda_1\\
-4\sqrt{(1-\lambda_1)(2-\lambda'_1-\lambda_1)}\le0.
\label{equiv cond}
\end{multline}
One can easily check that the left hand side of Eq.~(\ref{equiv cond}) vanishes if $\lambda'_1=1$, it is a convex function of~$\lambda_1$ and~$\lambda_1'$ and it is upper bounded by $\max\{3(\sfr73-\mu),0\}$ for any $0\le \lambda_1\le 1$, $0\le\lambda'_1\le 1$, thus the condition ${\mathsf c}^t({\mathsf M}-{\mathsf M}^{(0)}){\mathsf c}\le0$ holds if $\mu\ge 7/3$. But this implies that either the maximum value of the largest eigenvalue of ${\mathsf M}$ satisfies $\mu < 7/3$, or it is attained when $\lambda'_1=1$, that is, $\mu=\mu^{(0)}$.
A lower bound to $\mu$ is easily obtained by considering ${\mathsf M}$ for $L=2$, $\lambda'_1=1$ and $\lambda_1=0$:
\begin{eqnarray}
\widetilde{\mathsf M}:={\mathsf M}\bigg|_{\mbox{\raisebox{-.4em}{$\two{\lambda'_1=1}{\lambda_1=0}$}}}=
\begin{pmatrix}
{11\over6}&{\sqrt3\over2}&0\\[.5em]
{\sqrt3\over2}&{7\over4}&{1\over\sqrt2}\\[.5em]
0&{1\over\sqrt2}&1
\end{pmatrix}.\\
\nonumber
\end{eqnarray}
A lower bound to its eigenvalue, $\tilde\mu$, is given by ${\mathsf c}^t\widetilde{\mathsf M}{\mathsf c}$, for, e.g.,  ${\mathsf c}^t=(1,1,1)/\sqrt3$. It yields $\tilde\mu=2.58$, hence $7/3\le 2.58\le \tilde\mu\le\mu$, and we conclude that the maximum of ${\mathscr B}$ (i.e., of the largest eigenvalue of ${\mathsf M}$) is attained by~$\lambda'_1=1$.


It only remains to show that the optimal value of $\lambda_1$ is zero. To this end, let again ${\mathsf M}$ stand for the original matrix in Eq.~(\ref{M entries}) now with $\lambda_L=\lambda_{L-1}=\cdots=\lambda_2=0$ and~$\lambda'_1=1$. The matrix $\mathsf M$ is symmetric, tridiagonal and has positive off diagonal entries. 
The eigenvalues of such matrices are known to be simple and, hence, expected to be differentiable functions of their entries. By differentiating the eigenvalue equation ${\mathsf M}{\mathsf c}=\mu{\mathsf c}$ with respect to $\lambda_1$, where $\mathsf c$ is a normalized eigenvector (${\mathsf c}^t{\mathsf c}=1$), we obtain:
\begin{equation}
\mu'={\mathsf c}^t {\mathsf M}'{\mathsf c}, \quad \text{with}\ 0\leq\lambda_1<1.
\label{mu' M'}
\end{equation}
Here, the prime denotes the derivative with respect to $\lambda_1$, and we have used that the normalization condition implies ${\mathsf c}^t{\mathsf c}'=0$. One can readily see that
\begin{equation}
{\mathsf M}'=
\begin{pmatrix}
0&{1\over4\sqrt3}&0\\[.5em]
{1\over4\sqrt3}&-{3\over4}&-{1\over2\sqrt{2-2\lambda_1}}\\[.5em]
0&-{1\over2\sqrt{2-2\lambda_1}}&0
\end{pmatrix},
\end{equation}
where only the non vanishing entries, $l=2,1,0$, are displayed.
We now use again Eq.~(\ref{link c1 c0}), along with 
\begin{equation}
{\mathsf M}_{11} c_1+ {\mathsf M}_{10} c_{0}+{\mathsf M}_{21} c_{2} =\mu c_1 ,
\end{equation}
to obtain
\begin{align}
c_0&={\sqrt{1-\lambda_1}\over\sqrt2( \mu-1)}c_1,\\
c_2&={(3\mu-1)\lambda_1+4\mu^2-11\mu+5\over(\mu-1)(\lambda_1+6)}\sqrt3 c_1.
\end{align}
Eq.~(\ref{mu' M'}) becomes
\begin{equation}
\mu'= {(3\mu-1)\lambda_1+8\mu^2-40\mu+16\over4(\mu-1)(\lambda_1+6)}c_1^2<0,
\end{equation}
where we have used that $0\le\lambda_1\le1$ and $7/3\le\mu\le 4$. Hence $\mu$ is a decreasing function of $\lambda_1$ proving that the optimal value of this parameter is $\lambda_1=0$.


Finally, for $L=1$, we have
\begin{equation}
{\mathsf M}=
\begin{pmatrix}
1+{3\over4}\lambda'_1-{3\over4}\lambda_1&\sqrt{1-{1\over2}\lambda'_1-{1\over2}\lambda_1}\\
\sqrt{1-{1\over2}\lambda'_1-{1\over2}\lambda_1}&1
\end{pmatrix},
\end{equation}
and one can check explicitly that the statement of Theorem~\ref{theor 2} also holds true. This completes the proof~$\blacksquare$

\subsubsection{Proof of Theorem~\ref{theor 5}}

The proof closely follows the steps outlined in Sections~III and V of~\cite{gendra2013optimal}. The largest eigenvalue of~${\mathsf M}$ can be computed as the maximum value of ${\mathsf c}^t{\mathsf M}{\mathsf c}=\sum_{l,l'=0}^L{\mathsf M}_{ll'}c_lc_{l'}$, with $\sum_{l=0}^L{\mathsf c}_l^2=1$. For large $L$, we can approximate $l/L$ by a continuous variable $x\in[0,1]$ and~$c_l$ by a continuous function, $y(x)$, as $c_l\sim \sqrt\epsilon\, y(x)$, where $\epsilon=L^{-1}$. Likewise $c_{l-1}\sim \sqrt\epsilon\, y(x-\epsilon)
$,
\begin{equation}
{\mathsf M}_{ll}\sim{4x+3\epsilon\over 2x+2\epsilon},\quad
{\mathsf M}_{ll-1}\sim\sqrt{2x-\epsilon\over 2x}.
\end{equation}
Expanding in powers of $\epsilon$ and using the Euler-Maclaurin summation formula we have $\mu\sim 4-(2/L^2) S+o(L^{-2})$, where
\begin{equation}
S=\int_{0}^1\left[{y'^2\over2}+L{y^2\over2x}-{11y^2\over32x^2}-{\omega^2\over2}\left(y^2-1\right)\right] dx,
\label{new S}
\end{equation}
and to eliminate boundary terms we imposed the conditions~$y(0)=y(1)=0$. In this expression $\omega^2$ is a Lagrange multiplier enforcing the normalization constrain on $y$, i.e., $\int_0^1y^2 dx=1$. We can view~$S$ as an action and derive the ``equation of motion"  of $y$ from it:
\begin{equation}
-y''+L{y\over x}-{11y\over16x^2}=\omega^2y,\quad 0\le x\le 1.
\label{S-L 1}
\end{equation}
This equation determines the minimum of $S$, and hence the largest value of $\mu$. The extrema of $S$ are readily seen to be given by $\omega^2/2$, hence, 
$\mu_{\rm max}=4-\omega_{\rm min}^2/L^2$, where $\omega_{\rm min}$ is the smallest eigenvalue of the eigenvalue problem in Eq.~(\ref{S-L 1}) with boundary conditions $y(0)=y(1)=0$. 

While the eigenfunctions of this problem can be given in terms of Whittaker functions, for the purpose of computing the lowest orders in the asymptotic expansion of~$\mu_{\rm max}$ collected in Eq.~(\ref{expr theor 5}) it proves helpful to make a few approximations, which significantly simplify the resolution. Firstly, we drop the third term in Eq.~(\ref{S-L 1}) as it is subdominant compared to the second term.  Secondly, we expand $1/x$ (second term) in powers of $1-x$, as the eigenfunction $y$ is squeezed towards $x=1$ for large values of $L$ to minimize the action in Eq.~(\ref{new S}), hence, to first order we have $1/x\approx 2-x$. As a result, Eq.~(\ref{S-L 1}) is replaced~by
\begin{equation}
-y''+L(2-x)y=\omega^2y,\quad -\infty\le x\le 1,
\label{S-L 2}
\end{equation}
with boundary conditions $y(1)=\lim_{x\to-\infty}y(x)=0$.
Note that we have extended the domain of $y$ to the semi-infinity interval $(-\infty,1]$ and relaxed the boundary conditions. This procedure introduces exponentially decaying corrections to the asymptotic expansion of $\mu$, which we choose to neglect. The solution to Eq.~(\ref{S-L 2}) is given by
\begin{align}
y&=c\operatorname{Ai}\big(L^{1/3}-L^{1/3}x+\gamma_1\big),\label{sol eigenfunc}\\
 \omega^2&=L-L^{2/3}\gamma_1.
 \label{sol eigenval}
\end{align}
Here $\operatorname{Ai}$ is the Airy function of the first kind, the constant $\gamma_1\approx -2.3381$ is its first zero, and $c$ is an arbitrary constant which is fixed by normalization to be $c=L^{1/6}/\operatorname{Ai}'(\gamma_1)$, where $\operatorname{Ai}'$ is the first derivative of Airy function $\operatorname{Ai}$, and we have used the result $\int \operatorname{Ai}^2(x)dx=x\operatorname{Ai}^2(x)-\operatorname{Ai}'\!^2(x)$.

The next correction to~(\ref{sol eigenval}) is most conveniently computed by applying standard perturbation theory to the second order term in the expansion of $1/x$ [see Eq.~(\ref{S-L 1})] around $x=1$,  namely, to the ``perturbation" term~$(1-x)^2$. One has
\begin{multline}
L\!\!\int_{-\infty}^1\!\!(1\!-\!x)^2 y^2 dx=
c^2\!\! \int_{\gamma_1}^\infty\!(t\!-\!\gamma_1)^2\! \operatorname{Ai}^2(t)dt =\\
{8\over15}\gamma_1^2\operatorname{Ai}'\!^2(\gamma_1) c^2
={8\over15}L^{1/3}\gamma_1^2,
\end{multline}
where we have used again the expression of the primitive of $\operatorname{Ai}^2$ and integration by parts. Collecting the various contributions we readily obtain Eq.~(\ref{expr theor 5})~$\blacksquare$

\section{The optimal measure-and-operate strategy}\label{app:MO}

In Section~\ref{app:char MO}, we prove that the optimal measure-and-operate (MO) strategy can be realized by estimating the rotation applied to the atom and subsequently applying the estimated rotation to the target spin. The optimization of the MO fidelity is presented in Section~\ref{app:max MO}, and in Section~\ref{app:asymp MO}, we derive the asymptotic expansion of the MO fidelity.

\subsection{Characterizing the potentially optimal measure-and-operate channel}\label{app:char MO}
In line with the previous sections, we reexpress the fidelity $F^{\rm MO}(g)$ from Eq.~(\ref{main-MOfid}) of the main text as:
\begin{equation}
F_{\rm MO}(g) = \frac{1}{3} + \frac{2}{3} F_{\rm MO}^{(\rm e)}(g),
\label{FMO_e}
\end{equation}
where $F_{\rm MO}^{\rm (e)}(g)$ represents the entanglement fidelity~\cite{horodecki1999general} of the 
MO strategy and is defined as:
\begin{equation}
F_{\rm MO}^{\rm (e)}(g) = \frac{1}{4}\sum_k \bra{\phi_g}P_k\ket{\phi_g}\bb U_g|C_k|U_g\kk.
\label{FMO_initial}
\end{equation}
For the sake of brevity, we will refer to $F_{\rm MO}^{\rm (e)}(g)$ simply as ``fidelity" if there is no ambiguity or confusion.

Our goal now is to maximize $F_{\rm MO}^{\rm (e)}(g)$ in the worst-case scenario across all rotations $g$ applied to the sensor. 
To this end we begin by identifying the measurements and channels that can potentially lead to optimality. As discussed in the main text, the first crucial observation is that the channel ${\cal C}_k$ (with Choi operator $C_k$) can be chosen to be unitary without loss of generality. It follows that Eq.~(\ref{FMO_initial}) can be written as
\begin{equation}
F_{\rm MO}^{\rm (e)}(g) = \frac{1}{4}\sum_k \bra{\phi_g}P_k\ket{\phi_g}\, \left|\Tr\left[U^\dagger_g U_k\right]\right|^2,
\label{channel_unitary}
\end{equation}
where $U_k$ is the unitary transformation produced by the channel ${\cal C}_k$.

The next crucial observation states that the optimal MO strategy involves estimating the unknown rotation applied to the sensor and subsequently applying the estimated rotation to the target spin. This conclusion is a direct consequence of Theorem~\ref{theor mo yin 1} below. 
In line with the convention used in previous sections, we will denote the worst-case fidelity, $\inf_g F_{\rm MO}^{\rm (e)}(g)$, simply as $F^{\rm (e)}_{\rm MS}$, without argument.

\begin{theorem}\label{theor mo yin 1}
In order to maximize the worst-case entanglement fidelity $F^{\rm (e)}_{\rm MS}$, we can make the following assumptions without any loss of generality:
\begin{enumerate}
\item The measurement $(P_k)$ in Eq.~(\ref{channel_unitary}) is a rotationally covariant positive operator-valued measure (POVM) with a rank-one seed operator.
\item The unitary transformation $U_k$ applied to the target spin is the qubit rotation $U_{\hat g}$, where $\hat g$ represents the outcome of the covariant measurement.
\end{enumerate}
\end{theorem}
In other words, one can express the fidelity in Eq.~(\ref{channel_unitary})~as
\begin{equation}
F_{\rm MO}^{\rm (e)}(g) = \frac{1}{4}\int d\hat g \bra{\phi_g}P_{\hat g}\ket{\phi_g}\, \left|\Tr\left[U^\dagger_g U_{\hat g}\right]\right|^2,
\label{channel_unitary 2}
\end{equation}
with $P_{\hat g}=U_{\hat g}\Omega U^\dagger_{\hat g}$, where $\Omega=|\omega\>\<\omega|$ and $\int d\hat g P_{\hat g}=\id_{\rm S}$. The MO strategies satisfying conditions~1 and~2 in Theorem~\ref{theor mo yin 1} will be referred to as ``covariant strategies".


\medskip

\noindent{\bf Proof.} Let us define the average entanglement fidelity over all rotations as
\begin{equation}
\big\< F^{\rm (e)}_{\rm MO}\big\>=\int dg F^{\rm (e)}_{\rm MO}(g).
\label{<F>}
\end{equation}
We will prove the theorem in two steps. (i)~We will show that Theorem~\ref{theor mo yin 1} holds for the average fidelity~$\big\< F^{\rm (e)}_{\rm MO}\big\>$. (ii)~We will establish the equality between the maximum worst-case fidelity and the maximum average fidelity, namely that
$F^{\rm (e)}_{\rm MO,\,max} = \big\< F^{\rm (e)}_{\rm MO}\big\>_{\rm max}$.

\medskip

Consider the following chain of equalities: 
\begin{align}
	\nonumber \big\<F^{\rm (e)}_{\rm MO}\big\> &= \dfrac{1}{4} \sum_k \int d g \, \<\phi_g| P_k |\phi_g\> \, \Big| \Tr[U_g^\dagger U_k] \Big|^2\\
	\nonumber &= \dfrac{1}{4} \sum_k \int d g \, \<\phi_g| U_k U_k^\dagger P_k U_k U_k^\dagger |\phi_g\> \, \Big| \Tr[U_g^\dagger U_k] \Big|^2\\
	\nonumber &= \dfrac{1}{4} \int d g' \, \<\phi_{g'}|  \Big[\sum_k U_k^\dagger \, P_k \, U_k \Big] |\phi_{g'}\> \, \Big| \Tr[U_{g'}^\dagger] \Big|^2\\
	\nonumber &= \dfrac{1}{4} \int d g' \, \<\phi_{g'}| \, \Omega \, |\phi_{g'}\> \, \Big| \Tr[U_{g'}^\dagger] \Big|^2\\
	&= \dfrac{1}{4}  \int d \hat{g} \, \<\phi_h| \, P_{\hat{g}} \, |\phi_h\> \, \Big| \Tr[U_{h}^\dagger U_{\hat{g}}] \Big|^2 \, .
	\label{chain}
\end{align}
The resulting equation holds for any rotation $h$, and we have performed the following manipulations along the way: a change of variables $g\mapsto g'$ implicitly defined by the relation $U^\dagger_kU_g=U_{g'}$, 
applied the invariance of the Haar measure, introduced the definitions $\Omega := \sum_k U_k^\dagger \, P_k \, U_k$ and $P_{\hat{g}} := U_{\hat{g}} \, \Omega \, U_{\hat{g}}^\dagger$, and finally, in the bottom line, changed variables again, $g'={\hat g}^{-1}h$.

It is evident that $\Omega$ and $P_{\hat{g}}$ are non-negative operators, and one can readily verify that $\int d\hat{g} \, P_{\hat{g}} = \id_{\rm S}$, proving that the operators $P_{\hat g}$ define a POVM.  Since the fidelity is a linear function of $\Omega$, it follows that the optimal seed $\Omega$ can be chosen to be rank-one.

The bottom line of Eq. (\ref{chain}) represents the fidelity $F^{\rm (e)}_{\rm MO}(h)$ of a covariant strategy (as described in Theorem~\ref{theor mo yin 1}), for {\em any} rotation~$h$. By integrating over all such rotations, we establish that for any MO strategy, including the optimal one, there exists a covariant strategy that attains the same average fidelity. This proves the validity of statement (i).

Since Eq.~(\ref{chain}) holds for any rotation~$h$, it also implies the following relations:
\begin{equation}
\big\< F^{\rm (e)}_{\rm MO}\big\> = \inf_h F^{\rm (e)}_{\rm MO}(h) = F^{\rm (e)}_{\rm MO} \le F^{\rm (e)}_{\rm MO,\,max}.
\end{equation}
By optimizing the resulting bound over all MO strategies, we obtain the inequality $\big\< F^{\rm (e)}_{\rm MO}\big\>_{\rm max} \le F^{\rm (e)}_{\rm MO,\,max}$. 

On the other hand, we have the obvious inequality $\inf_h F^{\rm (e)}_{\rm MO}(h) \le F^{\rm (e)}_{\rm MO}(g)$, which holds for any rotation $g$ and any MO strategy. Integrating over $g$, this implies that $\inf_h F^{\rm (e)}_{\rm MO}(h) \le \big\< F^{\rm (e)}_{\rm MO}\big\>$. Therefore, we obtain the reverse inequality $F^{\rm (e)}_{\rm MO,\,max} \le \big\< F^{\rm (e)}_{\rm MO}\big\>_{\rm max}$. Hence, statement (ii) is proven~$\blacksquare$
\smallskip

\subsection{Maximizing the MO fidelity}\label{app:max MO}
\subsubsection{The optimal sensor state and POVM}
Let us recall that the state of the sensor, which is a hydrogen atom, has the form shown in Eq.~(\ref{c's}). In general, the unit vectors $|\phi_l\rangle$ can be expressed as:
\begin{equation}
|\phi_l\rangle = \sum_{m=-l}^{l} a^l_m |l,m\rangle , \quad \sum_{m=-l}^l |a^l_m|^2=1.
\label{phi & a}
\end{equation}
Similarly, the state $|\omega\rangle$ that defines the seed of the POVM as $\Omega=|\omega\rangle\langle\omega|$ has the form:
\begin{equation}
|\omega\rangle = \sum_{l=0}^L \sqrt{2l+1}|\omega_l\rangle,
\end{equation}
where $|\omega_l\rangle$ are also unit vectors in the spin-$l$ representation of ${\sf SU}(2)$, which are analogous to the vectors $|\phi_l\rangle$.

With this notation and recalling Theorem~\ref{theor mo yin 1}, we can rewrite Eq.~(\ref{channel_unitary}) as:
\begin{multline}
F^{\rm (e)}_{\rm MO} = 
	\dfrac{1}{4}\! +\! \dfrac{1}{4} \int d g \, |\<\phi_g| \omega\>|^2 \, \Tr[U_g^{(1)}]= \\
	\dfrac{1}{4}\! +\! \dfrac{1}{4} \sum_{l,l'=0}^L \!\!\dfrac{\sqrt{(2l\!+\!1)(2l^{'}\!+\!1)}}{3} \, \<\omega_l \tilde{\omega}_{l'}| \id^{(1)} |\phi_l \tilde{\phi}_{l'}\> \bar c_l c_{l'},
\end{multline}
where $U^{(1)}_g$ is the spin-$1$ irreducible representation of ${\sf SU}(2)$ and we have used the identity $\Tr{}^2[U_g]=1+\Tr[U^{(1)}_g]$ between the characters of the spin-$1/2$ and spin-$1$ irreps. We also recall the tilde notation, $|\tilde\phi\>={\rm e}^{-i\pi L_y}|\overline\phi\>$, introduced in the main text, where the bar stands for complex conjugation. Hereafter, the summation limits will be understood to ease the notation.

Hence, maximizing the MO fidelity is equivalent to maximizing $\operatorname{maxeigv}({\sf N})$ over all unit vectors $|\phi_l\>$ and $|\omega_l\>$, where ${\sf N}$ is the $(L\!+\!1)\!\times\!(L\!+\!1)$ 
matrix defined~by
\begin{equation}\label{matrix N}
	{\sf N}_{ll'} = \dfrac{\sqrt{(2l+1)(2l^{'}+1)}}{3} \<\omega_{l}\widetilde{\omega}_{l'}| \id^{(1)} |\phi_{l}\widetilde{\phi}_{l'}\> \, .
\end{equation}
To this end, we will first upper bound $\mu:=\operatorname{maxeigv}({\sf N})$ and then show that the bound is attainable, as we did in Section~\ref{app:optimality} for the general strategy.

A closer look to $\<\omega_{l}\tilde{\omega}_{l'}| \id^{(1)} |\phi_{l}\tilde{\phi}_{l'}\>$ reveals that this expression vanishes if $|l-l'| > 1$. Therefore the matrix $\sf N$ is tridiagonal. Furthermore, it is manifestly symmetric and non-negative (i.e., it has non-negative entries). This last property simplifies the optimization of $\mu$ significantly. The following theorem provides attainable bounds for the entries of $\sf N$ and states that these bounds are simultaneously achieved by choosing $|\phi_l\> = |\omega_l\> = |l,l\>$ for all $l\le L$. It is evident that the non-negativity of $\sf N$ guarantees that this choice maximizes its largest eigenvalue~$\mu$.

\begin{theorem}\label{theor mo yin 2}
The entries of the $(L+1)\times(L+1)$ matrix~$\sf N$ defined in Eq.~(\ref{matrix N}) can be bounded as follows:
\begin{equation}
{\sf N}_{ll}\le{l\over l+1}, \quad {\sf N}_{l\,l-1}\le \sqrt{2l-1\over2l+1},
\label{theor mo yin 2 expr}
\end{equation}
for all admissible values of $l$. These bounds are simultaneously attained when $|\phi_l\> = |\omega_l\> = |l,l\>$ for all $l\le L$.
\end{theorem}

Eq.~(\ref{main-FMO max}) in the main text directly follows from this theorem. Note that in the main text, the symbol $\sf N$ has a different meaning. It represents the matrix $\sf N$ defined in~(\ref{matrix N}) corresponding to the optimal states $|\phi_l\>$ and~$ |\omega_l\> $.

\subsubsection{Proof of Theorem~\ref{theor mo yin 2}}

{\em Diagonal entries.} We begin by expressing $\id^{(1)}|\phi_{l}\tilde{\phi}_{l}\>$ in the standard basis of the spin-$1$ irrep of ${\sf SU}(2)$. Using the Clebsch-Gordan coefficients, we obtain the following specific expression:
\begin{multline}
	\id^{(1)}|\phi_{l}\tilde{\phi}_{l}\>={(-1)^l\over \sqrt R}\Bigg[\!\left(\!\sum_m m\,a^l_m \bar{a}^l_{m}\!\right)\!|1,0\>\\
	\kern-3.5em -\left(\!\sum_m \!{\sqrt{(l\!+\!m)(l\!-\!m\!+\!1)}\;a^l_m \bar{a}^l_{m-1}}\!\right)\!{|1,1\>\over\sqrt2}  \\
	+\left(\!\sum_m\! {\sqrt{(l\!-\!m)(l\!+\!m\!+\!1)}\;a^l_m \bar{a}^l_{m+1}}\!\right)\!{|1,-1\>\over\sqrt2}\Bigg].
	\end{multline}
where
$
R=l(2l+1)(l+1)/3
$.
The three sums in this equation can be easily identified as the expectation values of the angular momentum operator $L_z$ and the ladder operators $L_-=L_x-iL_y$ and $L_+=L_x+iL_y$ on the state $|\phi_l\>$. As a result, we can express this equation as
\begin{multline}
	\id^{(1)}|\phi_{l}\tilde{\phi}_{l}\>=
	 {(-1)^{l}\over \sqrt R}\;
	 \Bigg[\av{L_{{z}}}_{\phi_l}|1,0\>\\
	-\! \dfrac{\av{L_{{-}}}_{\phi_l} }{\sqrt{2}}|1,1\>\!+\! \dfrac{\av{L_{{+}}}_{\phi_l}}{\sqrt{2}}|1,-1\> \Bigg].\label{I^(1) J's}
\end{multline}
%
An analogous expression holds for the~$\id^{(1)}|\omega_{l}\tilde{\omega}_{l}\>$. 

By considering Eq.~(\ref{matrix N}), we can obtain an upper bound for the diagonal entries of $\sf N$ as follows:
%
\begin{equation}
	{\sf N}_{ll} 
	 = \dfrac{1}{l(l+1)} \av{{\bf L}}_{\omega_l}\! \cdot \av{{\bf L}}_{\phi_l}
	\le\dfrac{l}{l+1} \, .
\end{equation}
We leave to the reader to check that the bound is achieved when $|\phi_{l}\> = |\omega_{l}\> = |l,l\>$.

\medskip

{\em Off-diagonal entries.} The projection $\id^{(1)}|\phi_{l}\tilde{\phi}_{l+1}\>$ can be written in the standard basis of the spin-$1$ irrep~as
\begin{equation}
\id^{(1)}|\phi_{l}\tilde{\phi}_{l+1}\>=\sum_{m'=-1}^1 \alpha_{m'}|1,m'\>,
\end{equation}
where we used the Clebsch-Gordan coefficients to compute the components $\alpha_m$. These components have the form
\begin{equation}
\alpha_{m'}={(-1)^{l+m'}\over \sqrt R} \sum_{m}a^l_m\sqrt{r^{m'}_{m}}\, \bar{a}^{l+1}_{m-m'}\, ,
\end{equation}
where we have conveniently redefined the coefficient $R$ as follows: $R=2(2l+3)(2l+1)(l+1)/3$. The remaining coefficients can be expressed as:
\begin{align}
r^1_m&=(l-m+2)(l-m+1)\label{r1},\\[.2em]
r^0_m&=2(l-m+1)(l+m+1)\label{r0},\\[.2em]
r^{-1}_m&=(l+m+2)(l+m+1)\label{r-1} .
\end{align}
Applying Cauchy-Schwarz inequality we bound $|\alpha_{m'}|^2$ as follows:
\begin{equation}
|\alpha_{m'}|^2\le {\left(\sum_{m}|a^l_m|^2\right)\over R}\sum_m r^{m'}_m|a^{l+1}_{m-m'}|^2.
\end{equation}
Hence
\begin{align}
\left\|\id^{(1)}|\phi_l\tilde\phi_{l+1}\>\right\|^2&\le {1\over R}\sum_{mm'} r^{m'}_m|a^{l+1}_{m-m'}|^2\\
&={1\over R}\sum_{m}\left(\sum_{m'} r_{m+m'}^{m'}\right)|a_m^{l+1}|^2 ,
\end{align}
where we have used the normalization condition in~(\ref{phi & a}).
Substituting Eqs.~(\ref{r1})--(\ref{r-1}) in  the inner sum we find
\begin{equation}
\sum_{m'} \!r_{m+m'}^{m'}\!=\!2[(2l+1)(l+1)-m^2]\le2(2l+1)(l+1),
\end{equation}
and we obtain the simple bound
\begin{equation}
\label{off bound}
\left\|\id^{(1)}|\phi_l\tilde\phi_{l+1}\>\right\|\le \sqrt{{3\over2l+3}} ,
\end{equation}
which is achieved if $|\phi_{l}\> = |l,l\>$ for all $l\le L$. The analogous bound also holds for the norm of~$\id^{(1)}|\omega_l\tilde\omega_{l+1}\>$.

Recalling Eq.~(\ref{matrix N}), we can upper bound the off-diagonal terms by applying Cauchy-Schwarz inequality:
\begin{equation}
\<\omega_l\tilde\omega_{l'}|\id^{(1)}|\phi_l\tilde\phi_{l'}\>\le
\left\|\id^{(1)}|\omega_l\tilde\omega_{l'}\>\right\|\left\|\id^{(1)}|\phi_l\tilde\phi_{l'}\>\right\| ,
\end{equation}
and using the bound~(\ref{off bound}) for the two norms. This~gives us:
\begin{equation}
{\sf N}_{ll+1}\le\sqrt{2l+1\over2l+3},
\end{equation}
which is equivalent to the second inequality in~(\ref{theor mo yin 2 expr}). This completes the proof. 

\subsection{Asymptotic MO fidelity}\label{app:asymp MO}

In this section we outline the derivation of the asymptotic form of the MO fidelity given in Eq.~(\ref{main-FMO}) of the main text.
Our starting point is the following expression of the maximum MO fidelity, which is equivalent to the bound~(\ref{main-FMO max}) in the main text, as this bound is attainable:
%
\begin{equation}\label{optimization MO}
	F^{\rm MO}_{\rm max} = {1\over2} + {1\over6}\max_{{\mathsf c}}{\mathsf c}^t{\mathsf N}{\mathsf c} .
\end{equation}
As in previous sections
${\mathsf c}^t=(c_L,c_{L-1},\dots, c_1,c_0)$, where the superscript $t$ denotes the transposed vector, and ${\mathsf c}^t{\mathsf c}=1$. The matrix ${\mathsf N}$ is given below equation~(\ref{main-FMO max}) in the main text. It is defined by~(\ref{matrix N}) setting $|\phi_l\>$ and $|\omega_l\>$ to be the optimal vector~$|l,l\>$. Specifically, the entries of ${\mathsf N}$ are given by
\begin{align}
	\displaystyle {\mathsf N}_{ll}={l\over l+1},\quad
	{\mathsf N}_{ll-1}=\sqrt{{2l-1\over2l+1}}.
	\label{N0 entries}
\end{align}
%
It follows that the maximum MO fidelity $F^{\rm MO}_{\rm max}$ can be expressed in terms of the largest eigenvalue of ${\mathsf N}$, denoted as $\mu$ throughout this section. More precisely, we have $F^{\rm MO}_{\rm max} = 1/2 + \mu/6$.
\begin{theorem}\label{theor 10}
	The asymptotic behavior of the largest eigenvalue of the matrix $\mathsf N$, defined in Eq.~(\ref{N0 entries}), is given~by
	\begin{equation}
		\mu\sim 3-{2\over L}+{2^{2/3}\gamma_1\over L^{4/3}}-{{8\over15}2^{1/3}\gamma_1^2\over L^{5/3}}+O(L^{-2}),
		\label{expr theor 6}
	\end{equation}
	where $\gamma_1\approx -2.3381$ is the first zero of the Airy function of the first kind, $\operatorname{Ai}(x)$.
\end{theorem}
\noindent{\bf Proof.} We follow the same approach as in the proof of Theorem~\ref{theor 5}, to which we refer the reader for more details. For large values of $L$, the optimization problem~(\ref{optimization MO}) can be reformulated as an extremization of a functional problem. An ``action" $S$ is defined to determine the largest eigenvalue of~$\mathsf N$ through the relation \mbox{$\mu\!=\!3\!-\!(2/L^2)S\!+\!o(L^{-2})$}.  Its explicit form is given by
\begin{equation}
	S=\int_{0}^1\left[{y'^2\over2}+L{y^2\over x}-{7y^2\over8x^2}-{\omega^2\over2}\left(y^2-1\right)\right] dx
	\label{S MO}
\end{equation}
and gives rise to the following ``equation of motion":
\begin{equation}
	-y''+2L{y\over x}-{7y\over4x^2}=\omega^2y,\quad 0\le x\le 1.
	\label{S-L MO}
\end{equation}
Using the same approximations as in the proof of Theorem~\ref{theor 5}, the solution to the eigenvalue problem~(\ref{S-L MO}) can be expressed as follows:
\begin{align}
	y&=c\operatorname{Ai}\big((2L)^{1/3}-(2L)^{1/3}x+\gamma_1\big),\label{sol eigenfunc}\\
	\omega^2&=2L-(2L)^{2/3}\gamma_1.
	\label{sol eigenval MO}
\end{align}
The next order correction to Eq.~(\ref{sol eigenval MO}) can be computed using perturbation theory, resulting in:
\begin{equation}
	2L\!\!\int_{-\infty}^1\!\!(1\!-\!x)^2 y^2 dx
	={8\over15}(2L)^{1/3}\gamma_1^2 .
\end{equation}
Gathering the various pieces, we obtain the expression given in Eq.~(\ref{expr theor 6}). This completes the derivation. $\blacksquare$

%
%

\medskip

The desired asymptotic expression of the MO fidelity readily follows from Theorem~\ref{theor 10}. 

\section{Entanglement between sensor and memory}\label{app:Ent}

In this section, we present the optimal learning strategy and calculate the corresponding fidelity when the sensor S and a specific portion of the memory labeled as ${\rm R}$ are entangled. Remarkably, the optimal strategy is determined to be of the measure-and-operate type, with a maximum fidelity exhibiting Heisenberg scaling, as shown in Eq.~(\ref{main-Fheisenberg}) of the main text.

Let $|\Phi\rangle_{\rm SR}$ denote the initial state of system SR. Following the interaction between the sensor and the environment, the state $|\Phi\rangle_{\rm SR}$ transforms into $|\Phi_g\rangle := (V_g \otimes \id_{\rm R})|\Phi\rangle_{\rm SR}$ and is stored in the quantum memory of the machine.
 Subsequently, the actuator will make use of the state $|\Phi_g\rangle$ to transform the unknown spin state $|\psi\>$, with the goal of replicating the desired state $U_g|\psi\>$. As in Sections~\ref{app:optimality} and~\ref{app:MO}, the overall performance is assessed based on the (gate)  fidelity
\begin{equation}
	F^{*}(g)=\!\!\int \!d\psi   \<\psi|U_g^\dagger \, \map{C}^{*}(|\Phi_g\>\<\Phi_g|\otimes|\psi\>\<\psi|)\, U_g|\psi\> ,
\end{equation}
where
$\map{C}^{*}$ is a retrieving channel, representing the process just described. After performing the integration over the target spin state $|\psi\rangle$, we arrive at an equation analogous to~(\ref{Fe F relat}) and~(\ref{FMO_e}),  with the entanglement fidelity now expressed as
\begin{equation}
	\kern-.2em F^{*(\rm e)}(g)\! =\!{1\over4}  \bra{U_g}\!\left({\mathcal C}^{*}\!\!\otimes\!{\mathcal I}_{{\rm T}_{\rm in}}\right)\!(|\Phi_g\>\<\Phi_g|\!\otimes\!|I\kk\bb I|)\ket{U_g},
	\label{def ent fid}
\end{equation}
where ${\cal I}_{{\rm T}_{\rm in}}$ is the identity channel acting on the target system ${{\rm T}_{\rm in}}$.
We can now apply the results presented by Bisio et al. in~\cite{bisio2010optimal} to derive the following expression, valid for ${\sf SU}(2)$:
%
\begin{equation}
F^{*(\rm e)} ={1\over4} \, {}_{\rm SR}\!\<\tilde\Phi|\,\bra{\!\< \id\,}\tilde C^*|\tilde\Phi\>_{\rm SR}\, \ket{\id\,\>\!} .
\label{wc Fe ent}
\end{equation}
Some comments are in order.
Ref.~\cite{bisio2010optimal} addresses a similar learning problem to ours, but allows for $N$ uses of the unknown unitary $U_g$ instead of a single use of $V_g$. The connection with their results requires setting $N=2L$. Additionally, the average fidelity considered in~\cite{bisio2010optimal} coincides with the worst-case fidelity~(\ref{wc Fe ent}). This equivalence can be proved by following a similar line of argument as presented in the proof of Theorem~\ref{theor mo yin 1}. 
It is worth noting that the optimality of Eq. (\ref{wc Fe ent}) relies on the optimality of ``parallel storing", as discussed in Ref. \cite{bisio2010optimal} (Lemma~1). 

The optimization over Choi operators~$C^*$ and states $|\Phi\>_{\rm SR}$, which constitutes the final step of the derivation, can also be borrowed from Ref.~\cite{bisio2010optimal}. According to the relevant results, the maximum value of $F^{{\rm (e)}}$ is achieved with a MO strategy, and the corresponding Choi operator can be expressed in our notation as:
\begin{equation}
C^*=\int d  g\, P_{  g}\otimes |U_{  g}\kk\bb U_{  g}| .
\end{equation}
In this equation, the set of all operators $P_{g}$, $g\in{\sf SU}(2)$, defines a POVM on system SR. These operators are given by:
\begin{equation}
P_{  g}\!=\!|\omega_{  g}\>\<\omega_{  g}|,\
|\omega_{  g}\>\!=\!\sum_{l=0}^L\!\sqrt{2l\!+\!1}\, U^{(l)}_{  g}\!\otimes\!\id^{(l)}_{\rm R} |\id^{(l)}\kk,
\end{equation}
 where $|\id^{(l)}\kk=\sum_{l=0}^L |l,m\>_{\rm S}\otimes|l,m\>_{\rm R}$.
This POVM represents an optimal choice for both estimating a rotation and transmitting a reference frame.
The optimal sen\-sor-mem\-o\-ry state is~(see also Ref.~\cite{bagan2004dense})
\begin{equation}
|\Phi\>_{\rm SR}={2\over\sqrt{2L+3}}\sum_{l=0}^L {\sin\left[{(2l+1)\pi\over 2L+3}\right]\over \sqrt{2l+1}} |\id^{(l)}\kk .
\end{equation}
With these results, we can derive the largest entanglement fidelity to be:
\begin{equation}
F^{*{\rm (e)}}_{\rm max}={1\over2}+{1\over2}\cos\left({2\pi\over N+3}\right) .
\end{equation}
Consequently, we find that
\begin{multline}
F^*_{\rm max}={2\over3}+{1\over3}\cos\left(2\pi\over 2L+3\right)\\
=1-{\pi^2\over 6L^2} + O\left(L^{-3}\right).
\end{multline}
The second line holds for large values of the angular momentum $L$. This equation corresponds to Eq.~(\ref{main-Fheisenberg}) in the main text.

\vfill


\onecolumngrid

\bibliographystyle{apsrev4-1}

\bibliography{learningatom}